\RequirePackage{fix-cm}
\documentclass[smallextended]{svjour3}       % onecolumn (second format)
\smartqed  % flush right qed marks, e.g. at end of proof
\usepackage{graphicx}
\usepackage{amsmath}	% Advanced maths commands
\usepackage{amssymb}	% Extra maths symbols
\usepackage{wasysym}
\usepackage{color,xcolor}
\usepackage{amsmath}
\usepackage{hhline}

\usepackage{cancel}
\begin{document}

\title{Dynamical taxonomy of the coupled solar radiation pressure and oblateness problem and analytical deorbiting configurations
%\thanks{}
}
% Grants or other notes about the article that should go on the front
% page should be placed within the \thanks{} command in the title
% (and the %-sign in front of \thanks{} should be deleted)
%
% General acknowledgments should be placed at the end of the article.


\titlerunning{SRP resonances}        % if too long for running head

\author{Ioannis Gkolias         \and
		Elisa Maria Alessi  	\and
        Camilla Colombo %etc.
}


\institute{I. Gkolias \at
				Department of Aerospace Science and Technology,
				Politecnico di Milano, Via la Masa 34, 20156 Milan, Italy
              \email{ioannis.gkolias@polimi.it}           %  \\
           \and
           E. M. Alessi \at
           Istituto di Matematica Applicata e Tecnologie Informatiche  "Enrico Magenes",
           Consiglio Nazionale delle Ricerche, 
           Via Alfonso Corti 12, 20133 Milano, Italy\\
           Istituto di Fisica Applicata "Nello Carrara",
           Consiglio Nazionale delle Ricerche, 
           Via Madonna del Piano 10, 50019 Sesto Fiorentino (FI), Italy
           \email{elisamaria.alessi@cnr.it}
           \and
           C. Colombo \at
           		Department of Aerospace Science and Technology,
				Politecnico di Milano, Via la Masa 34, 20156 Milan, Italy
               \email{camilla.colombo@polimi.it}  
}

\date{Received: date / Accepted: date}

\maketitle

\begin{abstract}
Recent works demonstrated that the dynamics caused by the planetary oblateness coupled with the solar radiation pressure can be described through a model based on singly-averaged equations of motion. The coupled perturbations affect the evolution of the eccentricity, inclination and orientation of the orbit with respect to the Sun--Earth line. Resonant interactions lead to non-trivial orbital evolution that can be exploited in mission design. Moreover, the dynamics in the vicinity of each resonance can be analytically described by a resonant model that provides the location of the central and hyperbolic invariant manifolds which drive the phase space evolution. The classical tools of the dynamical systems theory can be applied to perform a preliminary mission analysis for practical applications. On this basis, in this work we provide a detailed derivation of the resonant dynamics, also in non-singular variables, and discuss its properties, by studying the main bifurcation phenomena associated to each resonance. Last, the analytical model will provide a simple analytical expression to obtain the area-to-mass ratio required for a satellite to deorbit from a given altitude in a feasible timescale.
\end{abstract}

\section{Introduction}\label{sec:intro}

The effect of the Solar Radiation Pressure (SRP) on Earth satellites was recognised since the first space flights. The orbital evolution of Vanguard I \cite{Musen1960b} and the Echo balloons \cite{Shapiro1960} was found to be significantly influenced by SRP, and singly-averaged equations were used to study the motion \cite{Musen1960a}. Treated as a perturbation problem, the singly-averaged contribution of SRP is integrable and analytical solutions can be obtained \cite{Mignard1984,Oyama2008,Scheeres2012}. Nevertheless, when coupled with the effect of Earth's oblateness $J_2$ the system becomes a 2.5 Degrees-of-Freedom (DoF). An analytical insight can be recovered when treating locally the semi-secular SRP resonances. Namely, the singly-averaged perturbing function can be decomposed in six distinct terms \cite{Kaula1962}, each of them dominating in a particular range of orbital elements. The dynamics arising by combining each of the harmonics with the secular evolution due to $J_2$ can be reduced to a 1 DoF \emph{resonant} model \cite{Krivov1997,LucCol2012,Alessi_MNRAS}.  

The derivation of the six different resonant models in the three-dimensional case and their effect on the long-term evolution of resident space objects has been recently discussed in the literature \cite{Alessi_MNRAS,ACR_CMDA2019}. In this work, we re-derive the resonant models in the Hamiltonian framework, providing also a non-singular representation of the resonant dynamics. We exploit the information of the analytical model in \cite{ACR_CMDA2019} to obtain further insight in the resonant structures. We employ the equations for computing the equilibria for each resonance and compute the number and their stability for each set of the dynamical parameters of the system. This allows us to construct bifurcations diagrams which give the main transitions in the phase space. Particular focus is given to the structure of the phase space about the vicinity of each SRP resonance. The effect of the engineering parameter, the spacecraft area-to-mass ratio, is also discussed.

Using the phase space portraits, useful information for mission design concepts is retrieved via the tools of dynamical system theory. In particular, a rigorous procedure to obtain deorbiting conditions along the resonances is presented. As for the planar case \cite{LucCol2012}, we show how the deorbiting can occur. 

The paper is organised in the following way: in Sec.~\ref{sec:SRPmodel} the basic force model is recovered in its singly-averaged formulation, in Sec.~\ref{sec:resonances} we discuss the analytical description in the vicinity of SRP resonances, in Sec.~\ref{sec:phasespace} we provide a phase space analysis based on the main bifurcations associated to each resonance, in Sec.~\ref{sec:deorbiting} we demonstrate the use of the models to analytically obtain feasible deorbiting configurations, and in Sec.~\ref{sec:concl} we present our conclusions.

\section{Model derivation}\label{sec:SRPmodel}

Let us assume that a spacecraft moves under the effect of a planet's gravitational monopole, the planetary oblateness and the solar radiation pressure. Moreover, we assume that the sun-rays are always perpendicular to the surface of the satellite (cannonball model), that the effect of the planetary albedo is negligible, that the solar flux is constant at $1~au~$\color{black} and that the satellite is entirely in sunlight. Under these assumptions, SRP is modelled as a constant force in the direction of the Earth-Sun line, and it can be derived from a potential function.
The dynamics of the satellite in a geocentric equatorial inertial frame can be modelled by the Hamiltonian
$$
\mathcal{H} = \mathcal{H}_{kep} + \mathcal{H}_{J_2} + \mathcal{H}_{SRP}.
$$ 
The Keplerian part $\mathcal{H}_{kep}$ reads
$$
\mathcal{H}_{kep} = \frac{v^2}{2} - \frac{\mu}{r},
$$
where $\mu$ is the gravitational parameter of the Earth, and $r$, $v$ are the geocentric distance and  velocity of the satellite. 

\begin{figure}
\centering
\includegraphics[width=0.9\columnwidth]{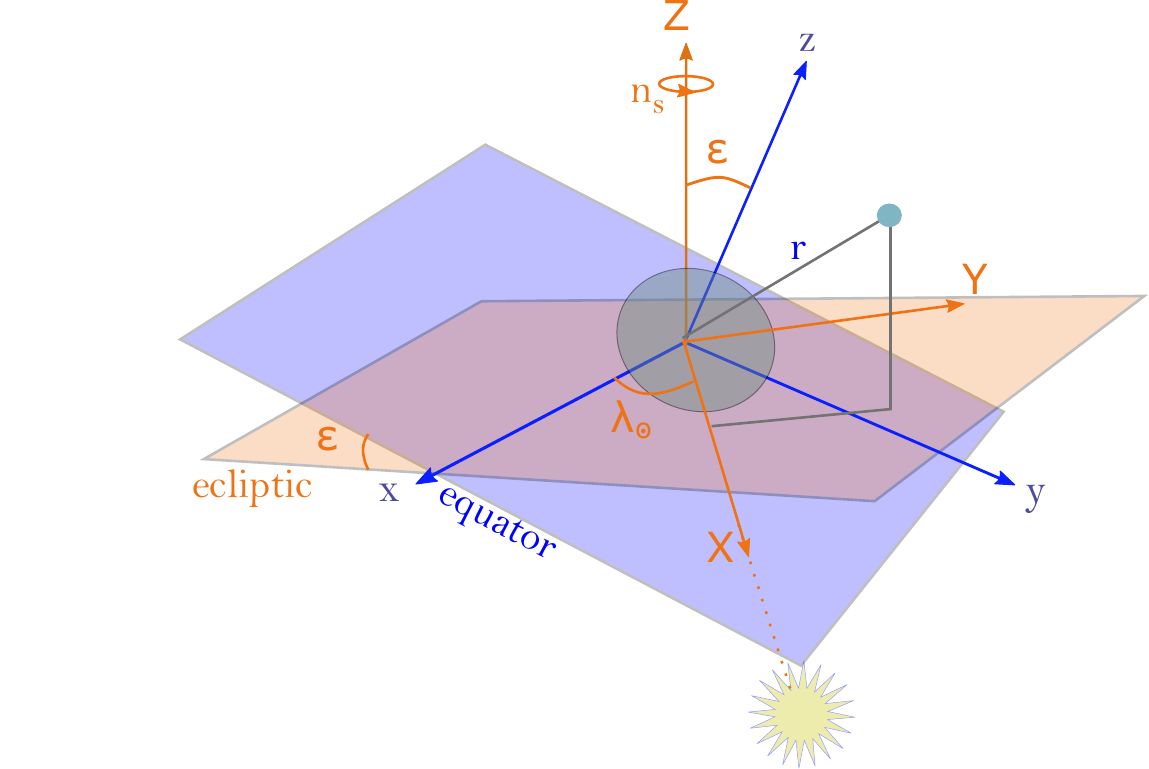}
\caption{Schematic description of an equatorial inertial and an ecliptic rotating reference frames. The solar radiation pressure is modelled as a constant force in the direction of the Sun.}
\label{fig:SRPframes}
\end{figure}

The Earth's oblateness effect is modelled as
$$
\mathcal{H}_{J_2} = \frac{ \mathcal{C}_{J_2}   \left(3 \sin^2\phi-1\right) }{2 r^3} , 
$$
where $\mathcal{C}_{J_2} = \mu R_{\oplus}^2 J_2$ with $J_2$ the oblateness parameter and $\phi$ the geographic latitude of the satellite. The sine of the latitude is expressed in terms of the orbital elements of the satellite via
$$ \sin \phi = \frac{z}{r},$$
and
$$
z =  (0,0,1) R_3( -\Omega) R_1(-i) R_3(-\lambda) (r,0,0)^{T},
$$
where $i$ is the inclination, $\Omega$ the Right Ascension of the Ascending Node (RAAN), $\lambda = \omega+f$ the argument of latitude\color{black}, $\omega$ the argument of perigee and $f$ the true anomaly of the satellite. The rotation matrices $R_1(u),R_3(u)$ are
\begin{equation}
R_1(u) = \left( 
\begin{array}{ccc} 
1 & 0 & 0 \\
0 & \cos u & \sin u \\
0 & -\sin u & \cos u
\end{array} \right) ,
\quad
R_3(u) = \left( 
\begin{array}{ccc} 
\cos u & \sin u & 0 \\
-\sin u & \cos u &  0 \\
0 & 0 & 1
\end{array} \right). 
\end{equation}
The solar radiation pressure contribution is given by
$$
\mathcal{H}_{SRP} = \mathcal{C}_{SRP} X,
$$
where  $ \mathcal{C}_{SRP} = P_\odot c_R \frac{A}{m}$, $P_\odot$ being the SRP constant at $1~AU$, $c_R$ the reflectivity coefficient and $\frac{A}{m}$ the area-to-mass ratio of the satellite. $X$ is the coordinate of the satellite in an Earth-centred system with the X-axis pointed towards the Sun (see Fig.~\ref{fig:SRPframes}): in terms of the orbital elements of the satellite it reads
$$
X =  (1,0,0) R_3( \lambda_\odot) R_1( \varepsilon)  R_3( -\Omega) R_1(-i) R_3(-\theta) (r,0,0)^{T},
$$
\color{black}
with $\lambda_\odot$ the ecliptic longitude of the Sun and $\varepsilon$ the obliquity of the ecliptic. 

Both $\mathcal{H}_{J_2}$ and $\mathcal{H}_{SRP}$ are considerable smaller than $\mathcal{H}_{kep}$ and can be treated as perturbations of the two-body problem
$$
\mathcal{H}  = \mathcal{H}_{kep}+\mathcal{H}_{per},
$$
where $\mathcal{H}_{per} = \mathcal{H}_{J_2}+ \mathcal{H}_{SRP}$. Employing a Hori-Deprit approach \cite{Hori1966,Deprit1969}, the homological equation up to the first order of the normalisation process reads:
$$
n \frac{\partial W_1}{\partial M} = \mathcal{H}_{per} - \bar{\mathcal{H}}_{per},
$$
where $n$ is the mean motion of the satellite, $W_1$ is the generating function and 
\begin{equation}
 \bar{\mathcal{H}}_{per} = \frac{1}{2 \pi} \int_0^{2 \pi} \mathcal{H}_{per} dM,
 \label{eq:avgint}
\end{equation}
is the first-order normalized Hamiltonian. The integrals in Eq.~(\ref{eq:avgint}) can be carried out in closed form for both the $J_2$ and SRP contributions. In the case of $J_2$ we use the differential relationship $dM = \frac{r^2}{a \sqrt{1-e^2}}df$ along with $r= a \frac{1-e^2}{1+e \cos{f}}$ to obtain the classical result
\begin{equation}
\bar{\mathcal{H}}_{J_2} =\frac{\mathcal{C}_{J_2} (3 c_i^2 -1)}{4 a^3 (1-e^2)^{3/2}}, 
\end{equation}   
where $c_i = \cos{i}$. For SRP we express the integral with respect to the eccentric anomaly $E$ using the relations $r \sin f = a \sqrt{1-e^2} \sin E$, $r \cos f = a(\cos E -e)$, $r=a(1-e \cos{E})$ and  $dM = \frac{r}{a} dE$. 

\noindent
The averaged model is then \cite{Krivov1997,LucCol2012,Alessi_MNRAS} 
\begin{equation}
\bar{\mathcal{H}}_{SRP} =  - \frac{3}{2}  a e  \mathcal{C}_{SRP}\sum_{j=1}^6 \mathcal{T}_j \cos{\psi_j}, 
\label{eq:hamSRP}
\end{equation}
with
\begin{equation}
\begin{array}{|c|c|c|}
\hline
j  &\mathcal{T}_j & \cos\psi_j  \\
\hline\hline
1 & \frac{1}{4} \left(c_{\varepsilon }+1\right) \left(c_i+1\right) & \cos{( \omega + \Omega   - \lambda_\odot)}  \\
2 & -\frac{1}{4} \left(c_{\varepsilon }+1\right) \left(c_i-1\right) & \cos{(- \omega + \Omega  - \lambda_\odot)}  \\
3 & \frac{1}{2} s_i s_{\varepsilon } & \cos{(\omega - \lambda_\odot)}    \\
4 & -\frac{1}{2}s_i s_{\varepsilon } & \cos{(\omega + \lambda_\odot)}    \\
5 & -\frac{1}{4} \left(c_{\varepsilon }-1\right) \left(c_i+1\right) & \cos{( \omega + \Omega + \lambda_\odot)} \\
6 & \frac{1}{4} \left(c_{\varepsilon }-1\right) \left(c_i-1\right) & \cos{(-\omega + \Omega  + \lambda_\odot)} \\
\hline
\end{array}
\label{eq:TjSRP}
\end{equation}
 \color{black}
where $c_i, s_i,c_\varepsilon, s_\varepsilon$ are the cosine and sine of the inclination $i$ and the obliquity of the ecliptic $\varepsilon$, respectively. The model consists of 6 distinct harmonics describing the semi-secular evolution of the system, all of them containing the ecliptic longitude of the Sun $\lambda_\odot$. A second averaging over the Sun's mean motion results in a null Hamiltonian, that is, the SRP perturbation does not give rise to secular effects. Moreover, we should mention that the Hamiltonian Eq.~(\ref{eq:hamSRP}) is integrable, and an analytical solution is obtained using an ecliptic rotating frame \cite{Mignard1984}. 

The singly-averaged model of the coupled Earth's oblateness and solar radiation pressure effects reads
$$
\mathcal{\bar{H}} = \bar{\mathcal{H}}_{J_2} + \bar{\mathcal{H}}_{SRP}.
$$
The Hamiltonian can be expressed in terms of the canonical Delaunay elements $(L,G,H,l,g,h)$such that
\begin{equation}
\begin{aligned}
L &= \sqrt{\mu a},  \\
G &= L \sqrt{1-e^2}, \\
H &= G \cos i , 
\end{aligned}
\quad
\begin{aligned}
l &= M, \\
g &= \omega, \\
h &= \Omega,
\end{aligned}
\end{equation}
to obtain
\begin{equation}
\mathcal{\bar{H}} =    \frac{\mathcal{C}_{J_2}  \mu^3(G^2 - 3 H^2)}{4 G^5 L^3} - \frac{3 \mathcal{C}_{SRP} \sqrt{1 - \frac{G^2}{L^2}} L^2}{2 \mu} \sum_{j=1}^6 \mathcal{T}_j \cos{\psi_j},
\label{eq:hamalldel}
\end{equation}
where for the coefficients $\mathcal{T}_j$ we use the relationships $c_i = H/G, s_i = \sqrt{1-\frac{H^2}{G^2}}$.

Due to the averaging process, Eq.~(\ref{eq:hamalldel}) does not depend on the mean anomaly $M=l$ and thus the Delaunay action $L$, as well as the semi-major axis, are constant. The system has two Degrees-of-Freedom (DoF) and an explicit time dependence through $\lambda_{\odot}(t) = \lambda_{\odot,0}+ n_s t$, where $n_s$ is the frequency corresponding to Earth's orbital period of 1 year. Considering an extended phase-space, a dummy action $I_s$ with frequency $n_s$ is added to the Hamiltonian 
$$
\mathcal{\bar{H}} =   \frac{\mathcal{C}_{J_2}  \mu^3(G^2 - 3 H^2)}{4 G^5 L^3} - \frac{3 \mathcal{C}_{SRP} \sqrt{1 - \frac{G^2}{L^2}} L^2}{2 \mu} \sum_{j=1}^6 \mathcal{T}_j \cos{\psi_j} + n_s I_s,
$$
which yields a three DoF autonomous system. 

\subsection{Semi-secular solar gravitational resonances}\label{sec:semisec}

As a side note, we recall that the second order singly-averaged Hamiltonian perturbative term corresponding to the solar gravitational attraction, say $\mathcal{\bar{H}}_{3bS}$, includes the same resonant arguments reported in (\ref{eq:TjSRP}). In particular, 
\begin{equation}\label{eq:Hj3b}
\begin{aligned}
\mathcal{\bar{H}}_{3bS}^1&=-\frac{15 a^2 e^2 \mu_S \left(c_{\varepsilon }+1\right)^2 \left(c_i+1\right)^2 \cos (2 \psi_ 1)}{128 a_S^3}, \\
\mathcal{\bar{H}}_{3bS}^2&=-\frac{15 a^2 e^2 \mu_S  \left(c_{\varepsilon }+1\right)^2 \left(c_i-1\right)^2 \cos (2 \psi_2)}{128 a_S^3},\\
\mathcal{\bar{H}}_{3bS}^3&=\frac{15 a^2 e^2 \mu_S   \left(c_{\varepsilon }^2-2 s_{\varepsilon }^2-1\right)s_i^2 \cos (2\psi_3)}{64 a_S^3}, \\
\mathcal{\bar{H}}_{3bS}^4&=\frac{15 a^2 e^2 \mu_S   \left(c_{\varepsilon }^2-2 s_{\varepsilon }^2-1\right)s_i^2 \cos (2\psi_4)}{64 a_S^3},\\
\mathcal{\bar{H}}_{3bS}^5&=-\frac{15 a^2 e^2 \mu_S  \left(c_{\varepsilon }-1\right)^2 \left(c_i+1\right)^2 \cos (2 \psi_5)}{128 a_S^3},\\
\mathcal{\bar{H}}_{3bS}^6&=-\frac{15 a^2 e^2 \mu_S  \left(c_{\varepsilon }-1\right)^2 \left(c_i-1\right)^2 \cos (2 \psi_6)}{128 a_S^3},
\end{aligned}
\end{equation}
where $\mu_S$ is the solar gravitational parameter and $a_S$ the geocentric semi-major axis of the Sun.

\noindent
In the following, we will assume that the range of $(a,e)$ and the value of area-to-mass ratio are such that the perturbation effects corresponding to $\mathcal{\bar{H}}_{3bS}$ are negligible. This is, for $j$ given, 
$$
\left|\frac{\mathcal{\bar{H}}_{SRP}^j}{\mathcal{\bar{H}}_{3bS}^j}\right|\gg 1.
$$
For the specific application presented at the end of the paper (Sec.~\ref{sec:deorbiting}), we will verify this hypothesis.

\section{Resonances}\label{sec:resonances}   

We are now interested in studying the resonance phenomena occurring in the coupled system. A resonance is occurring when the following commensurability holds:
\begin{equation}
n_1 \dot{g} + n_2 \dot{h} + n_3 \dot{\lambda}_{\odot} \approx 0
\label{eq:rescond} 
\end{equation}
where $\dot{g}=\partial \bar{\mathcal{H}}/\partial G$, $\dot{h}=\partial \bar{\mathcal{H}}/ \partial H$ and $\dot{\lambda}_{\odot} = \partial \bar{\mathcal{H}}/ \partial I_s$. The integer coefficients $n_1,n_2,n_3$ can take the values of the 6 combinations associated to the harmonics appearing in Eq.~(\ref{eq:TjSRP}). As a first approximation, the angular frequencies can be computed by taking into account only the $H_{J_2}$ part of the Hamiltonian, namely,
\begin{equation}
\begin{aligned}
\dot{g}_{J_2} &= \frac{\partial \bar{\mathcal{H}}}{\partial G} = -\frac{3 (G^2 - 5 H^2) \mathcal{C}_{J_2} \mu^3}{4 G^6 L^3} = -\frac{3}{2}\frac{J_2 R^2_{\oplus}n}{a^2(1-e^2)^2}\cos i, \\
\dot{h}_{J_2} &= \frac{\partial \bar{\mathcal{H}}}{\partial H} = -\frac{3 H \mathcal{C}_{J_2} \mu^3}{2 G^5 L^3} =  \frac{3}{4}\frac{J_2 R^2_{\oplus}n}{a^2(1-e^2)^2}\left(5\cos^2i-1\right), \\
\dot{\lambda}_{\odot} &= \frac{\partial \bar{\mathcal{H}}}{ \partial I_s} = n_s.
\end{aligned}
\label{eq:J2anglerates}
\end{equation}
Substituting Eq.~(\ref{eq:J2anglerates}) in Eq.~(\ref{eq:rescond}), we can derive an approximation of the loci in action-space of the center of each resonance \cite{Cook1962,Alessi_MNRAS}. In the vicinity of each resonance, the associate harmonic is slow and is dominating the dynamics. Assuming an isolated resonance approximation, the system can be described in the vicinity of the particular resonance by \color{black} the Hamiltonian
\begin{equation}
\mathcal{H}_{\psi_j} =  \frac{\mathcal{C}_{J_2}  \mu^3(G^2 - 3 H^2)}{4 G^5 L^3} - \frac{3 \mathcal{C}_{SRP} \sqrt{1 - \frac{G^2}{L^2}} L^2}{2 \mu} \mathcal{T}_j \cos{\psi_j} + n_s I_s.
\label{eq:reshamdel}
\end{equation}
 
For each resonance with argument $\psi_j$ \color{black} (with  $j = 1\ldots 6$) we introduce a resonant set of variables ($\Psi,\psi$) via a unimodular transformation of the Delaunay elements 
\begin{equation}
\begin{aligned}
\Psi &= \frac{G}{n_1}, \\
\Pi &= -n_2 G+n_1 H, \\
K &=  I_s - \frac{n_3 G}{ n_1}.
\end{aligned}
\quad
\begin{aligned}
\psi &= n_1 g + n_2 h + n_3 \lambda_{\odot}, \\
\pi &= \frac{h}{n_1}, \\
\kappa &= \lambda_{\odot}.
\end{aligned}
\end{equation}
which brings the Hamiltonian to a 1-DoF form of
$$
H_{\psi_j} = H_{\psi_j,J_2}(\Psi; L,\Pi) + H_{\psi_j,SRP}(\Psi,\psi; L,\Pi) + n_s (n_3 \Psi + K), 
$$
where the part associated with $J_2$ is
\begin{eqnarray}\label{eq:ham_j2}
\mathcal{H}_{\psi_j,J_2}(\Psi;L,\Pi)&=&\frac{C_{J_2} \mu^3 \left(n_1^4 \Psi^2 -3 (\Pi + n_1 n_2 \Psi)^2\right)}{ n_1^7 4 L^3 \Psi^5},
\end{eqnarray}
and the one due to SRP is
\begin{eqnarray}\label{eq:ham_SRP}
\mathcal{H}_{\psi_j,SRP}(\Psi,\psi;L,\Pi)&=&-\frac{3 C_{SRP}}{2\mu} L^2 \sqrt{1-\frac{n_1^2\Psi^2}{L^2}}  \mathcal{T}_j \cos\psi,
\end{eqnarray}
while the coefficients $\mathcal{T}_j$ are expressed in terms of the new variables using the equations $c_i= \frac{n_1n_2\Psi+\Pi}{n_2^2\Psi}$ and $s_i = \sqrt{1-c_i^2}$. 

Due to the resonant transformation both $\pi$ and $\kappa$ are ignorable, therefore, $\Pi$ and $K$ are constants and the term $n_s K$ can be dropped from the Hamiltonian. The action variable $\Pi$ is a resonant integral of the system; its value is dictated from the initial conditions and remains constant during the orbital evolution. Expressed in orbital elements
\begin{equation}\label{eq:Pi}
\Pi = \sqrt{\mu a} \sqrt{1-e^2} (-n_2  + n_1 \cos{i})
\end{equation}
represents coupled oscillations in the eccentricity and inclination of the orbit, in a similar fashion to the Lidov-Kozai constant in the third-body gravitational perturbations \cite{Lidov1962,Kozai1962}. Note that $\Pi$ corresponds to $\Lambda$ in \cite{ACR_CMDA2019}. The formulation provided here is equivalent to the one given in the past paper, but $n_1$ and $n_2$ are exchanged.

For a satellite close to a resonance with argument $\psi_j$ \color{black} with mean initial elements
$$(a_0,e_0,i_0,\omega_0,\Omega_0,\lambda_{\odot,0}),$$ 
the orbit evolution in the $(\Psi,\psi)$ plane, or equivalently in the $(e,\psi)$ plane if we substitute $\Psi = \sqrt{\mu a (1-e^2)}/n_1$, is given from the contour line of the implicit equation:
$$
H_{\psi_j}(\Psi,\psi;L,\Pi) = H_{\psi_j}(\Psi(a_0,e_0),\psi(\omega_0,\Omega_0,\lambda_{\odot,0});L,\Pi),
$$
with $L = L(a_0), \Pi = \Pi(a_0,e_0,i_0)$. Notice that $\Pi$ depends on the resonance $j$ through $n_1$ and $n_2$.

The equations of the motion related the resonant models $H_{\psi_j}$ are
\begin{equation}
\begin{aligned}
\frac{d \psi}{dt} = \frac{\partial H_{\psi_j}}{\partial \Psi},  \\
\frac{d \Psi}{dt} = - \frac{\partial H_{\psi_j}}{\partial \psi},
\end{aligned}
\end{equation}   
and the associated equilibria are given from the equations 
\begin{equation}
\begin{aligned}
\frac{d \psi}{dt} = 0, \\
\frac{d \Psi}{dt} = 0.
\end{aligned}
\end{equation}
Those stationary solutions of the resonant model represent periodic orbits of the full equations of motion. Their stability is determined from the eigenvalues of the Hessian of the Hamiltonian $\mathcal{H}_{\psi_j}$ computed at each equilibrium point.

In many applications the equations of motion with respect to the eccentricity are of interest, and thus, by applying the chain rule for the derivatives and using the relations $\Psi= G/n_1$ and $de/dG = - \frac{\sqrt{1-e^2}}{e \sqrt{\mu a}}$ we can derive the semi-canonical form:
\begin{align*}
\frac{de}{dt} = n_1 \frac{ \sqrt{1-e^2}}{e \sqrt{\mu a}} \frac{\partial H_{\psi_j}}{\partial \psi} = n_1 \mathcal{C}_{SRP} \frac{ \sqrt{1-e^2}}{na} \mathcal{T}_j \sin \psi \\
\frac{d\psi}{dt} = - n_1 \frac{ \sqrt{1-e^2}}{e \sqrt{\mu a}} \frac{\partial H_{\psi_j}}{\partial e}  = n_1 \dot{g}_{(J_2,\psi_j)} + n_2 \dot{h}_{(J_2,\psi_j)}  + n_3 n_s.
\end{align*}
and
\begin{equation}\label{eq:fullanglerates}
\begin{aligned}
\dot{g}_{(J_2,\psi_j)} = \dot g_{J_2} + \dot g_{\psi_j}, \\
\dot{h}_{(J_2,\psi_j)} = \dot h_{J_2} + \dot h_{\psi_j},
\end{aligned}
\end{equation}
where $\dot{g}_{J_2},\dot{h}_{J_2}$ are given in Eq.~(\ref{eq:J2anglerates}) and 
\begin{equation}
\begin{aligned}
\dot h_{\psi_j}&=\frac{\partial H_{\psi_j,SRP}}{\partial H} = C_{SRP}\frac{e}{na\sqrt{1-e^2}\sin i} \frac{\partial \mathcal{T}_{j}}{\partial i} \cos{\psi},\\
\dot g_{\psi_j}&= \frac{\partial H_{\psi_j,SRP}}{\partial G} = C_{SRP}\frac{\sqrt{1-e^2}}{nae} \mathcal{T}_{j}\cos{\psi}-\dot h_{\psi_j}\cos i,
\end{aligned}
\end{equation}
that is the model developed in \cite{ACR_CMDA2019}.

In all the above formulation the angles $g,h$ are not well defined when the eccentricity and/or the inclination are zero. To alleviate this burden we use the modified Delaunay variables in Eq.~(\ref{eq:reshamdel})
\begin{equation}
\begin{aligned}
\Lambda &= L,\\
P &= L-G, \\
Q &= G-H,
\end{aligned}
\quad
\begin{aligned}
\lambda &= l+g+h, \\
p &= -g-h, \\
q &= -h, 
\end{aligned}
\end{equation}
while a unimodular transformation allows to derive a new set of resonant variables ($\Sigma,\sigma$), similar to the ones introduced in \cite{Breiter2001} for the case of lunisolar resonances, namely,
\begin{equation}
\begin{aligned}
\Sigma &= \frac{P}{k_1},\\
\Phi &= -k_2 P + k_1 Q, \\
\Gamma &= I_s - \frac{k_3 P }{k_1},
\end{aligned}
\quad
\begin{aligned}
\sigma &= k_1 p + k_2 q + k_3 \lambda_{\odot},\\
\phi &= \frac{q}{k_1}, \\
\gamma &= \lambda_{\odot}. 
\end{aligned}
\end{equation}
where $k_1,k_2,k_3$ are integers appearing in combinations associated to each harmonic in Eq.~(\ref{eq:hamSRP})
\begin{equation}
\begin{array}{|c|c|c|}
\hline
j  & \cos{\psi_j} & (k_1,k_2,k_3) \\
\hline\hline
1 &  \cos{(- p  - \lambda_\odot)} & (\mp 1,0,\mp 1)\\
2 &  \cos{(-p + 2q - \lambda_\odot)} & (\mp 1,\pm 2,\mp 1) \\
3 &  \cos{(- p + q - \lambda_\odot)}  & (\mp 1,\pm 1,\mp1)  \\
4 & \cos{(-p + q + \lambda_\odot)} & (\mp 1, \pm 1,\pm 1) \\
5 &  \cos{( - p + \lambda_\odot)} & (\mp 1,0, \pm 1)\\
6 &  \cos{(-p + 2 q + \lambda_\odot) )} & (\mp 1,\pm 2,\pm 1)\\
\hline
\end{array}
\end{equation}
After a Taylor expansion in the resonant action $\Sigma$ and dropping the constant terms, the resonant Hamiltonian up to a truncation order $N$ takes the form
\begin{equation}
H_{\psi_j}^{(N)} = n_s k_3 \Sigma + C_{J_2} \sum_{n=1}^N c_n \Sigma^n + C_{SRP} \sum_{n=1}^N d_n \Sigma^{n-1/2} \cos{\sigma},
\label{eq:hamrestrunc}
\end{equation}   
where $c_n$, $d_n$ are constant coefficients depending on the dynamical parameters $(L,\Phi)$, the particular resonance $(k_1,k_2)$ and the physical parameters $(\varepsilon, \mu)$. We notice that up to third order ($N=3$) the resonant model is similar to the Extended Fundamental Model introduced in \cite{Breiter2001} and analyzed in \cite{Breiter2003}. In practice, a truncation order of $N \geq 4$ is required to retrieve the qualitative features of the phase space, while higher orders are used for more accurate computations. Actually, the phase-space exploration can be done with the non-truncated resonant Hamiltonian and we retain the complete dynamical portrait of the resonance, a fact that highlights the power of the closed-form averaging process. 

Finally, it is common to express the resonant Hamiltonian in terms of Poincar\'e variables
\begin{equation}
\begin{aligned}
X = \sqrt{2 \Sigma} \cos{\sigma}, \quad Y = \sqrt{2 \Sigma} \sin{\sigma},
\end{aligned}
\end{equation} 
with $x = \frac{X}{\sqrt{L}} \sim e \cos{\sigma}$ and $y = \frac{X}{\sqrt{L}} \sim e \sin{\sigma}$ providing a polar representation of the resonant domain, while the truncated Hamiltonian in Eq.~(\ref{eq:hamrestrunc}) takes a polynomial form in $x,y$ coordinates. 

\section{Bifurcation analysis}\label{sec:phasespace}    

From the equations written above it is possible to compute the equilibrium points associated with the dynamical system together with the corresponding stability. The isolated resonance system has two constants of motion (apart from $\mathcal{H}$): the semi-major is constant due to the averaging and the second integral $\Pi$ from Eq.~(\ref{eq:Pi}). The values of $\Pi$ can be labelled based on the inclination of the corresponding circular orbit $\Pi\equiv\Pi_{circ} = n_1 \cos{i_{circ}} \sqrt{\mu a} - n_2$. Given a set of values of the dynamical parameters ($a$, $i_{circ}$) and the engineering parameter $A/m$, a phase space is uniquely identified. In this phase space the number of equilibrium points and their stability can be defined. By varying the parameters and tracking the structural changes in the system, a bifurcation diagram can be computed. Elliptic and saddle points appear and disappear based on the classical bifurcation theory for 1-DOF systems.

\begin{figure}
\centering
\includegraphics[width=0.45\columnwidth]{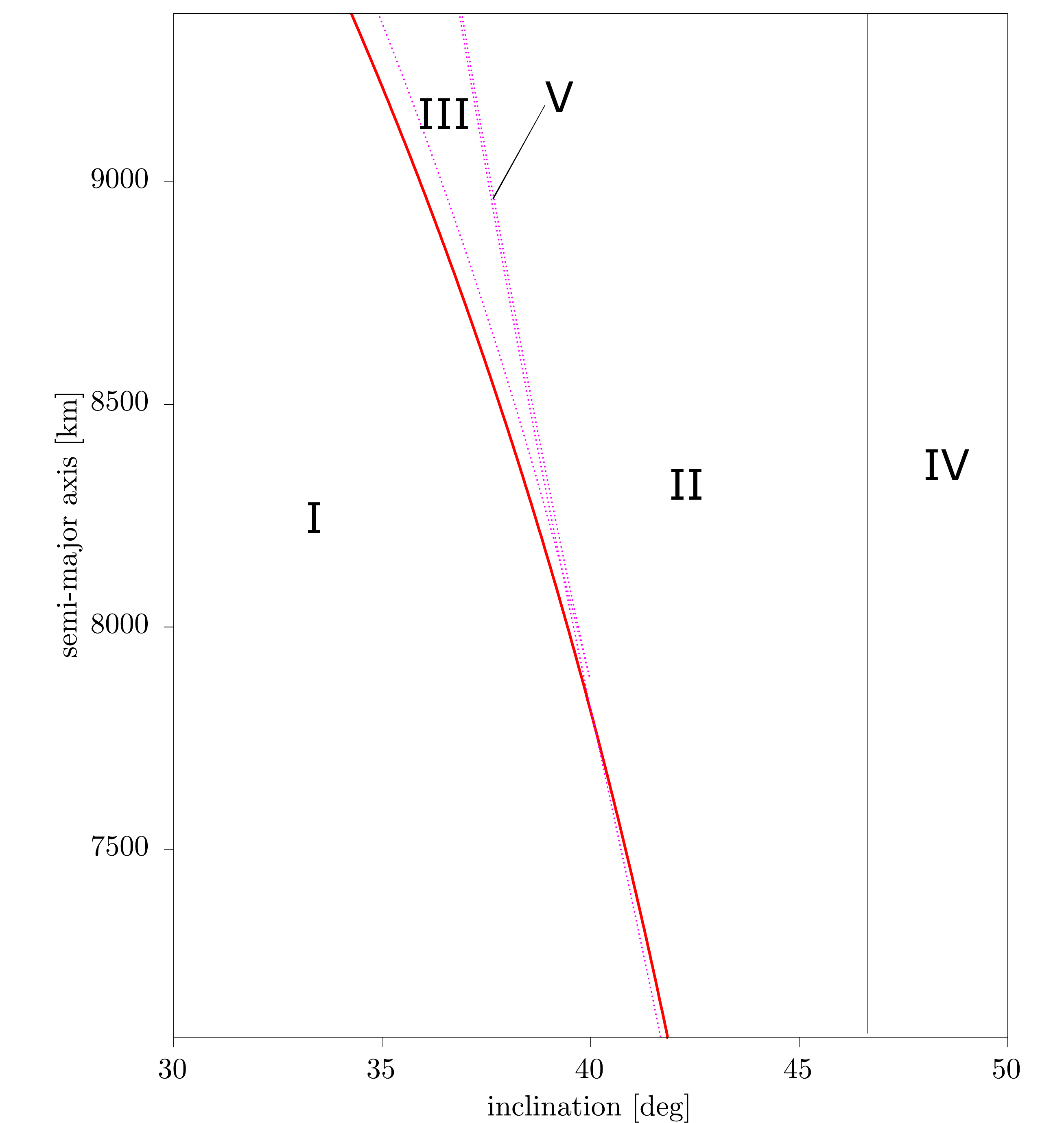}
\includegraphics[width=0.45\columnwidth]{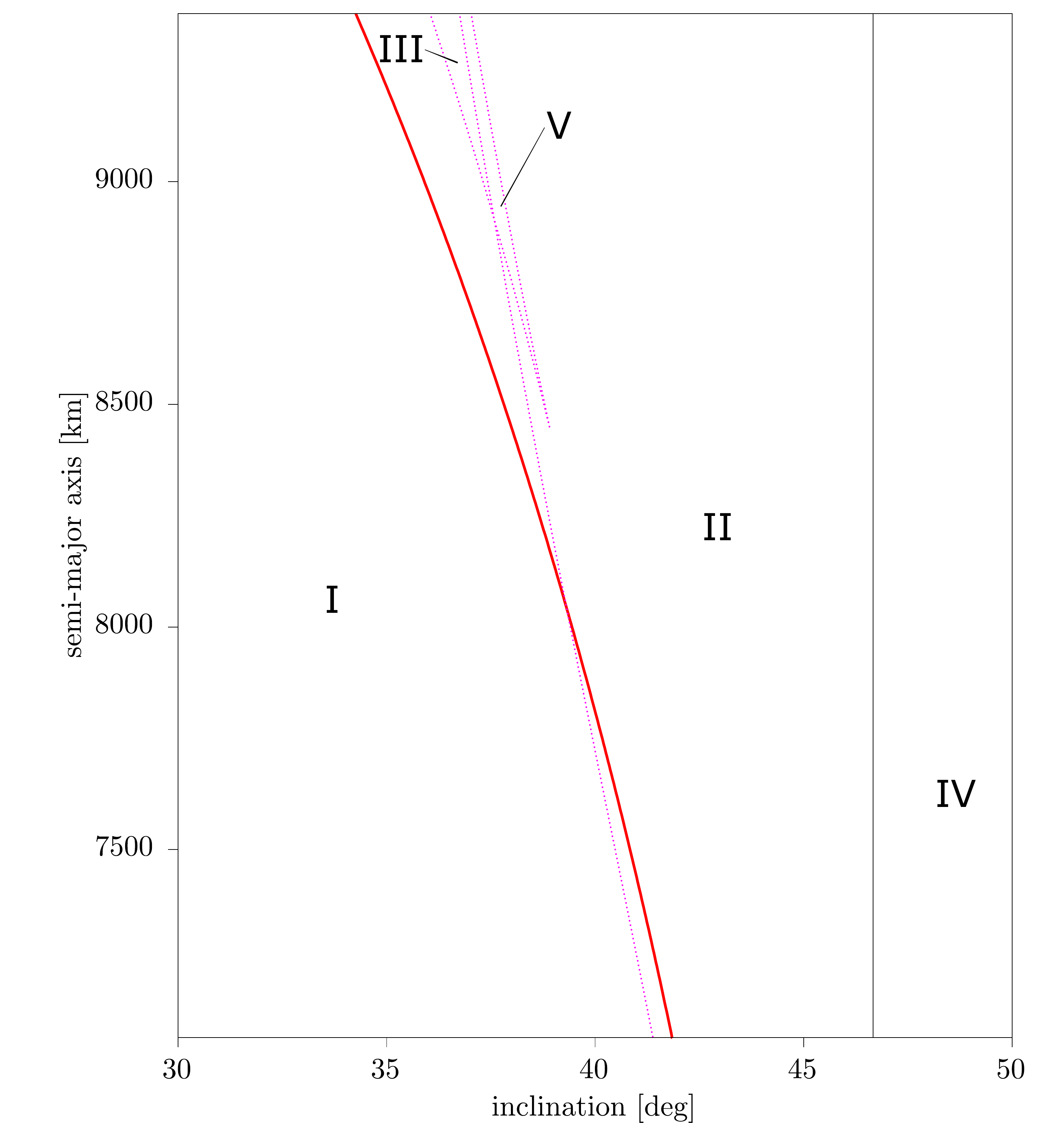}
\caption{The bifurcation diagram for the resonance with argument $\psi_1$. The total number of equilibria is presented in the dynamical parameter space $(a,i_{circ})$. Left: $A/m =1$ m$^2/$kg; right: $A/m =5$ m$^2/$kg. The red line corresponds to the resonant condition computed from solely the $J_2$ rate. Inclination stands for inclination of the circular orbit. For the type of equilibria in each region see Tab.~\ref{tab:res1eq}.}
\label{fig:bifplotpsi1}
\end{figure}

\begin{figure}[htbp!]
\centering
\includegraphics[width=0.6\columnwidth]{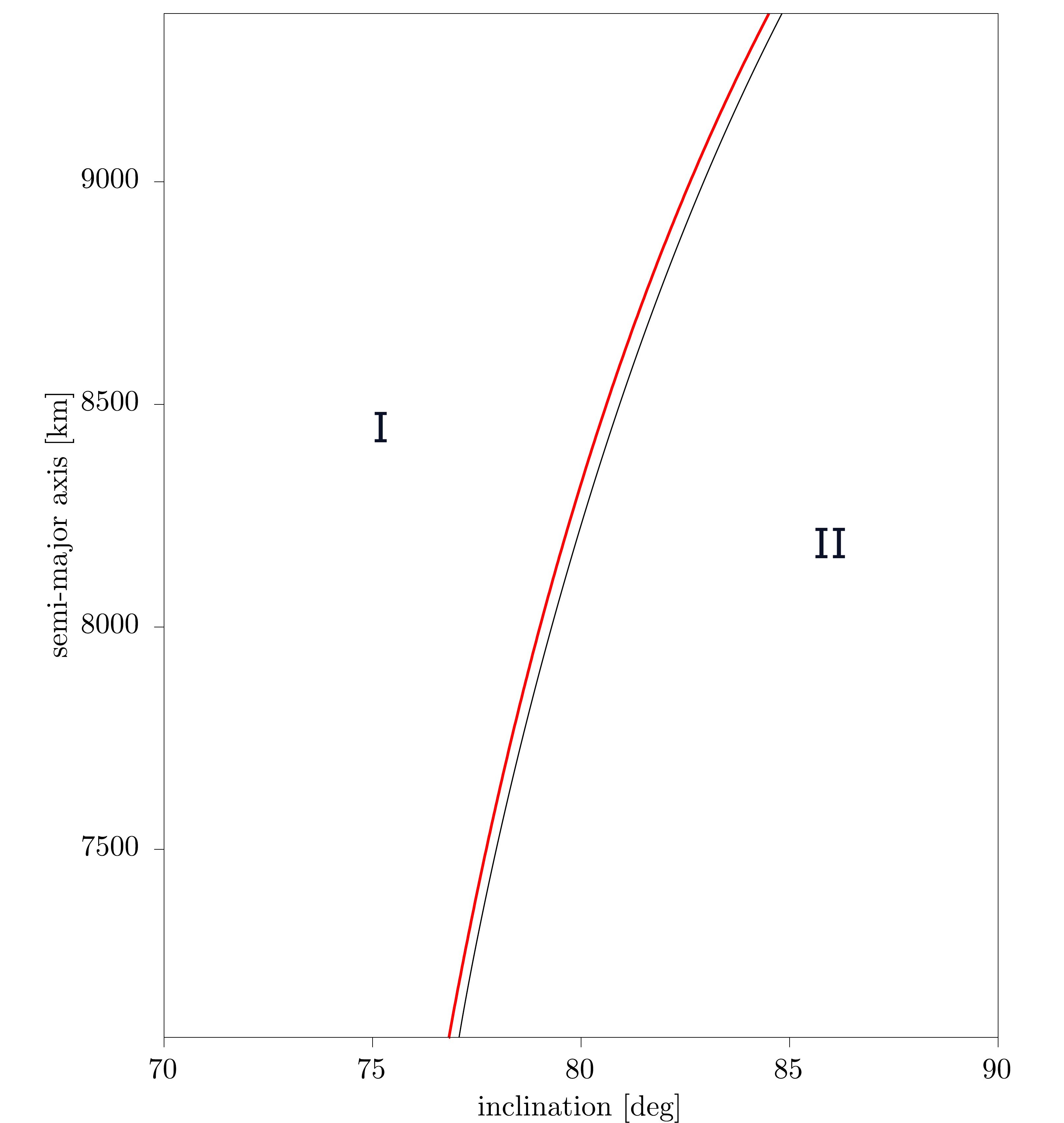}
\caption{The bifurcation diagram for the resonance with argument $\psi_2$ assuming $A/m =1$ m$^2/$kg. The total number of equilibria is presented in the dynamical parameter space $(a,i_{circ})$. The red line corresponds to the resonant condition computed from solely the $J_2$ rate. Inclination stands for inclination of the circular orbit. For the type of equilibria in each region see Tab.~\ref{tab:res2eq}.}
\label{fig:bifplotpsi2}
\end{figure}

For the two harmonics that dominate in the low Earth orbit region, namely, $\psi_1$ and $\psi_2$ \cite{ASRV_CMDA2018},  in Fig.~\ref{fig:bifplotpsi1} and in Fig.~\ref{fig:bifplotpsi2} the possible phase space structures are depicted in the range of semi-major axis $a\in[7000:9400]$ km. The black curves define the boundaries of the regions characterized by a different number of equilibrium points and corresponding stability. In red, the curves that would be obtained by considering only the oblateness effect for the rate of precession of $\omega$ and $\Omega$ are shown, that is, only  Eq.~(\ref{eq:J2anglerates}) instead of  Eq.~(\ref{eq:fullanglerates}). 

\begin{figure}
\centering
\includegraphics[width=0.35 \columnwidth]{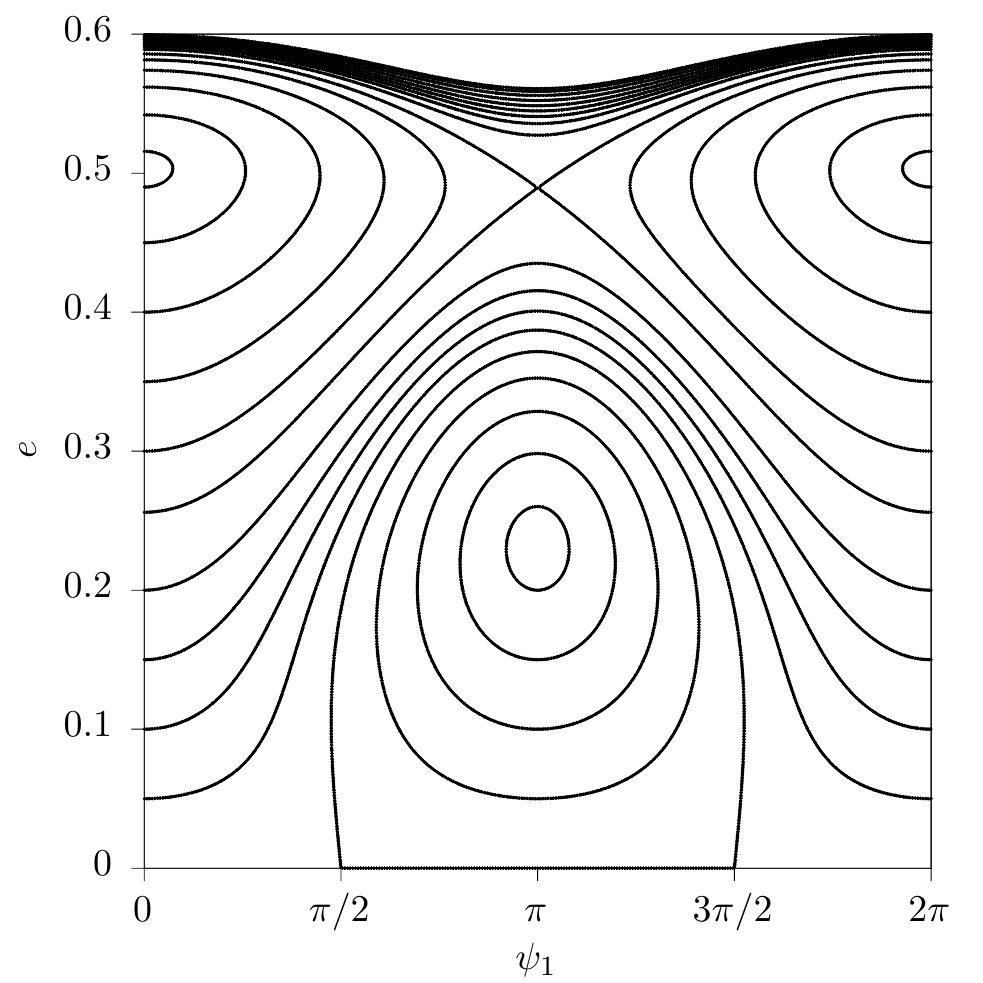}
\hspace{5pt}
\includegraphics[width=0.35 \columnwidth]{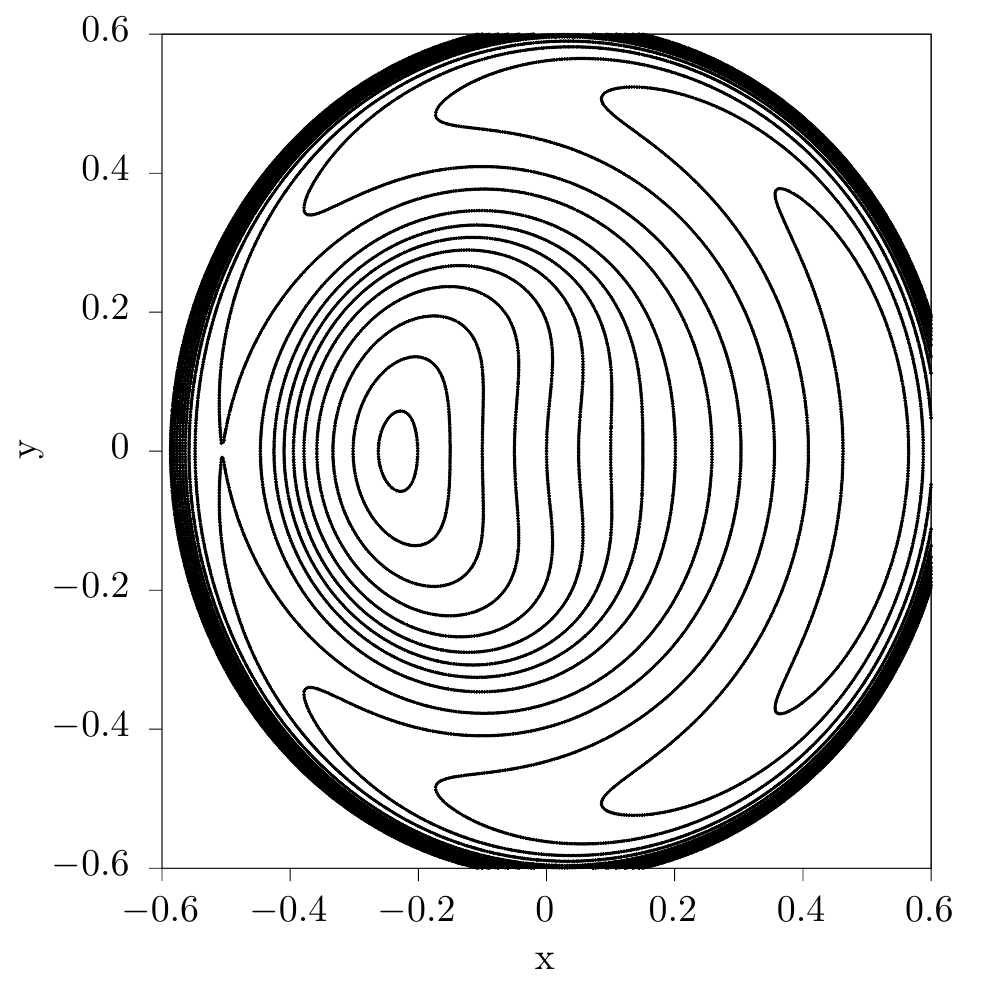}\\
\includegraphics[width=0.35 \columnwidth]{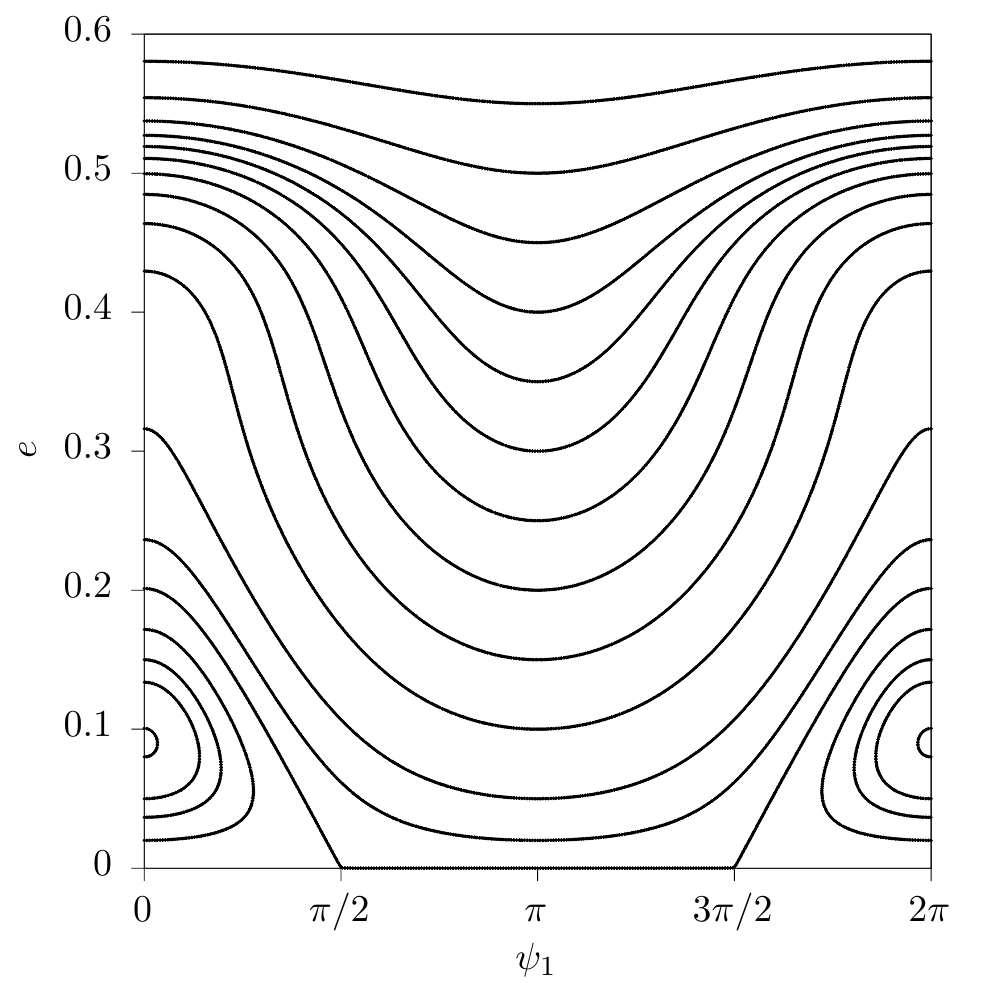}
\hspace{5pt}
\includegraphics[width=0.35 \columnwidth]{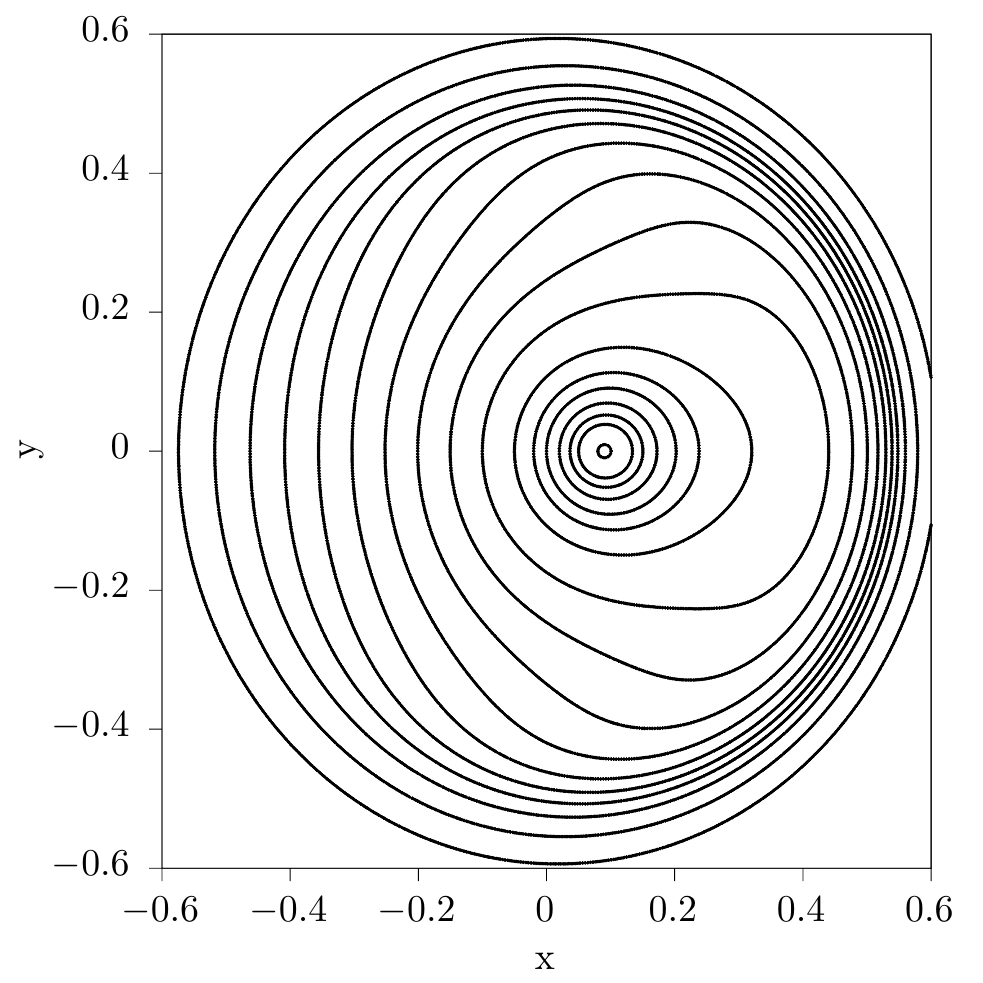}\\
\includegraphics[width=0.35 \columnwidth]{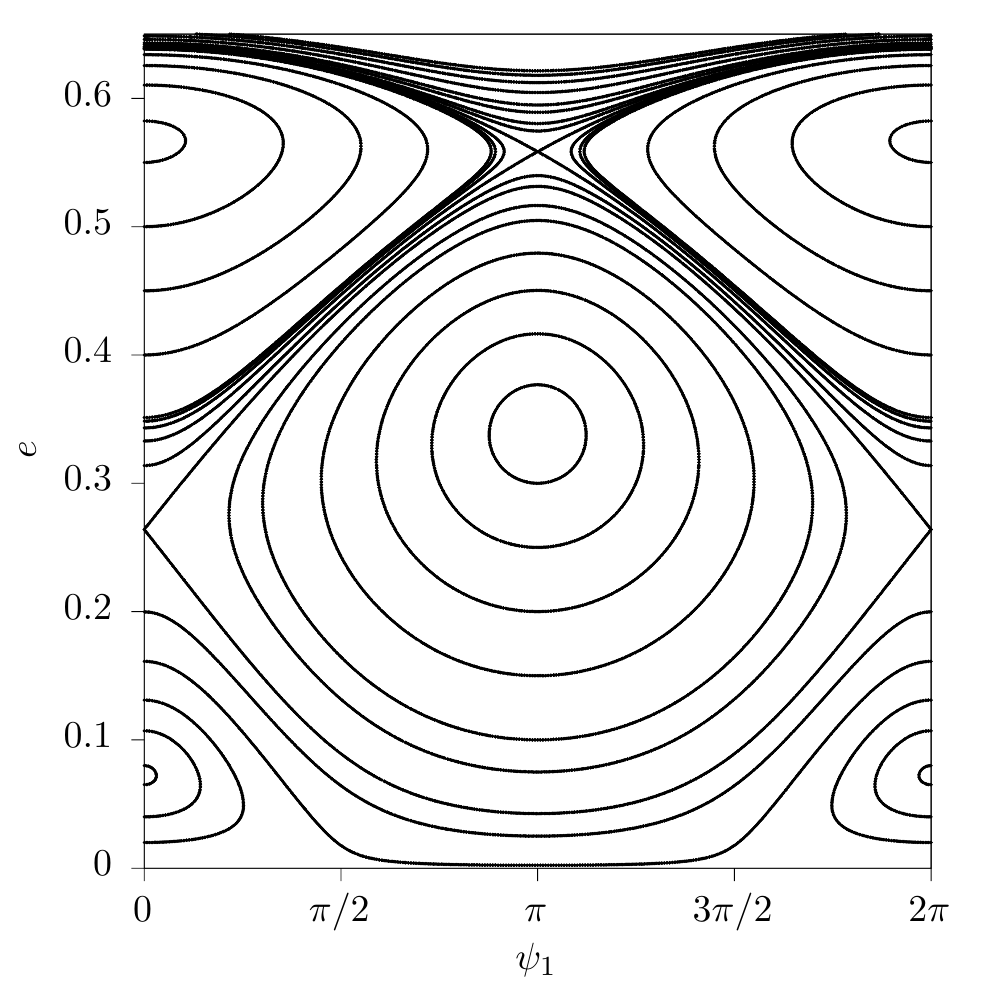}
\hspace{5pt}
\includegraphics[width=0.35 \columnwidth]{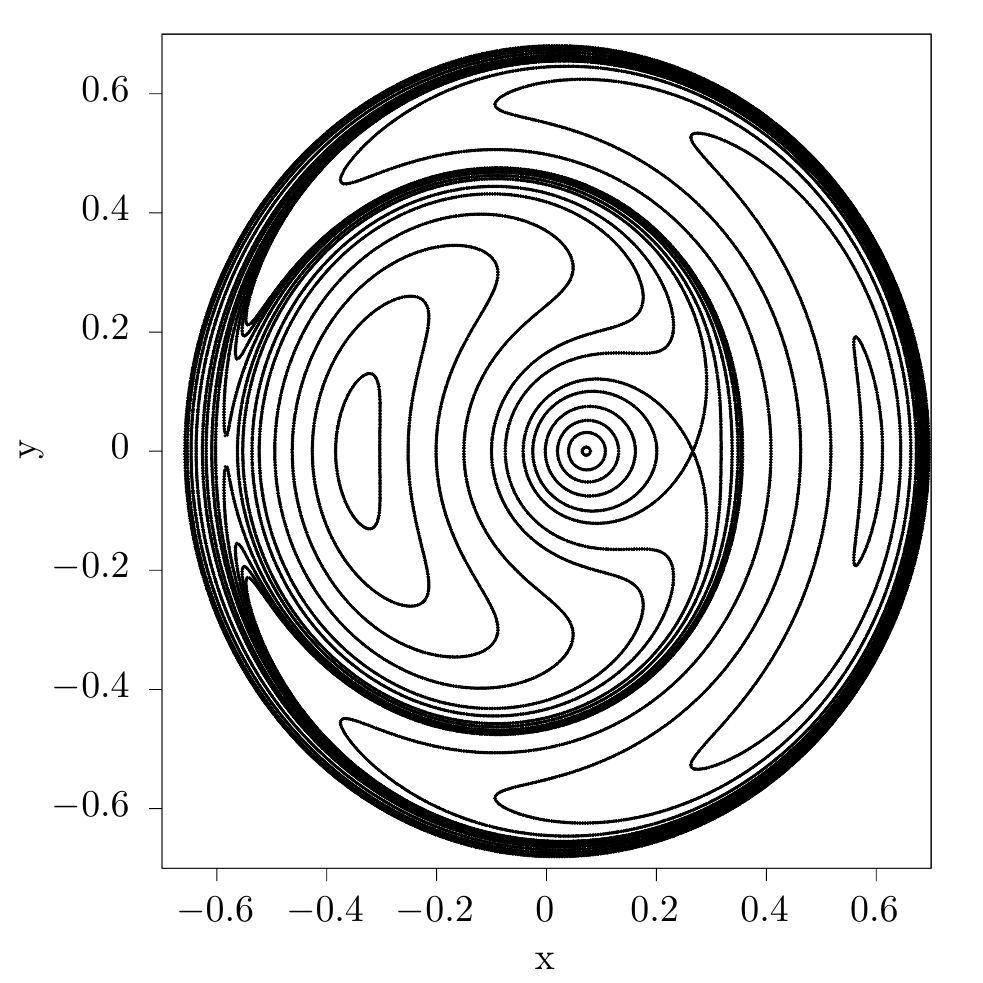}\\
\includegraphics[width=0.35 \columnwidth]{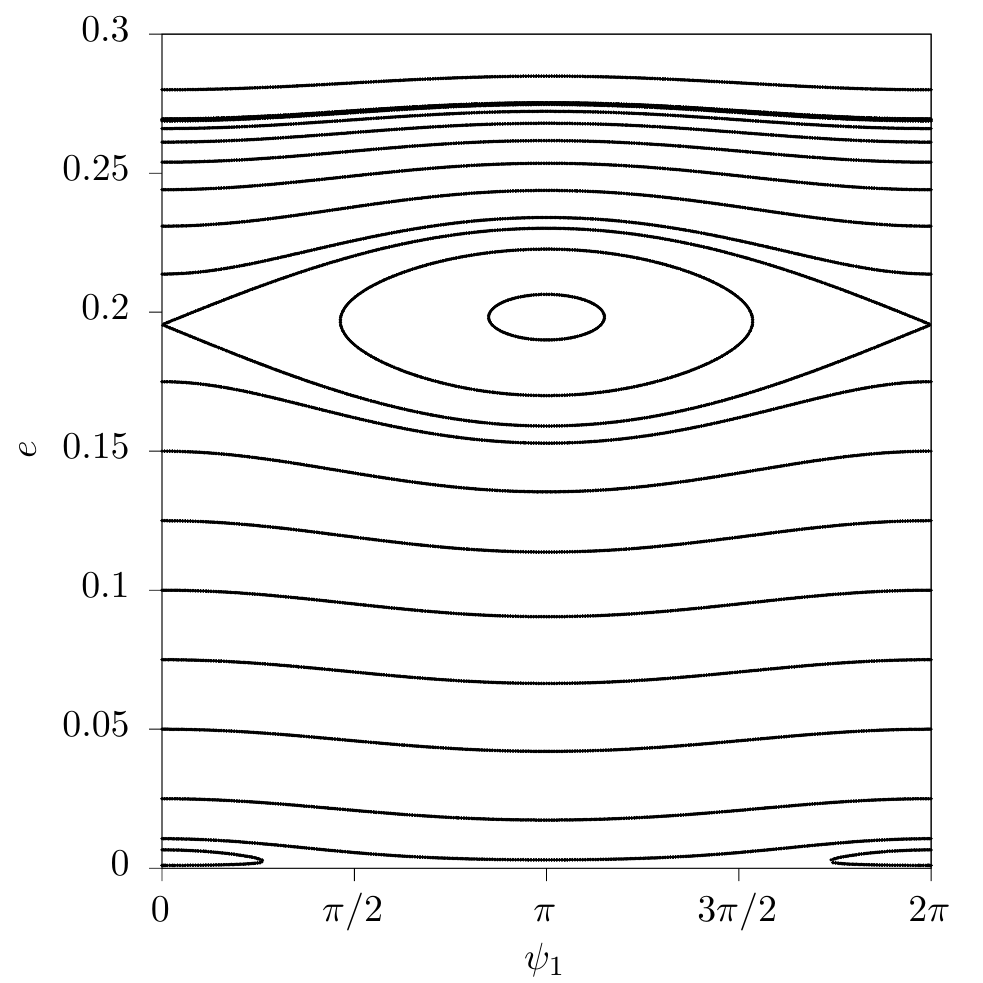}
\hspace{5pt}
\includegraphics[width=0.35 \columnwidth]{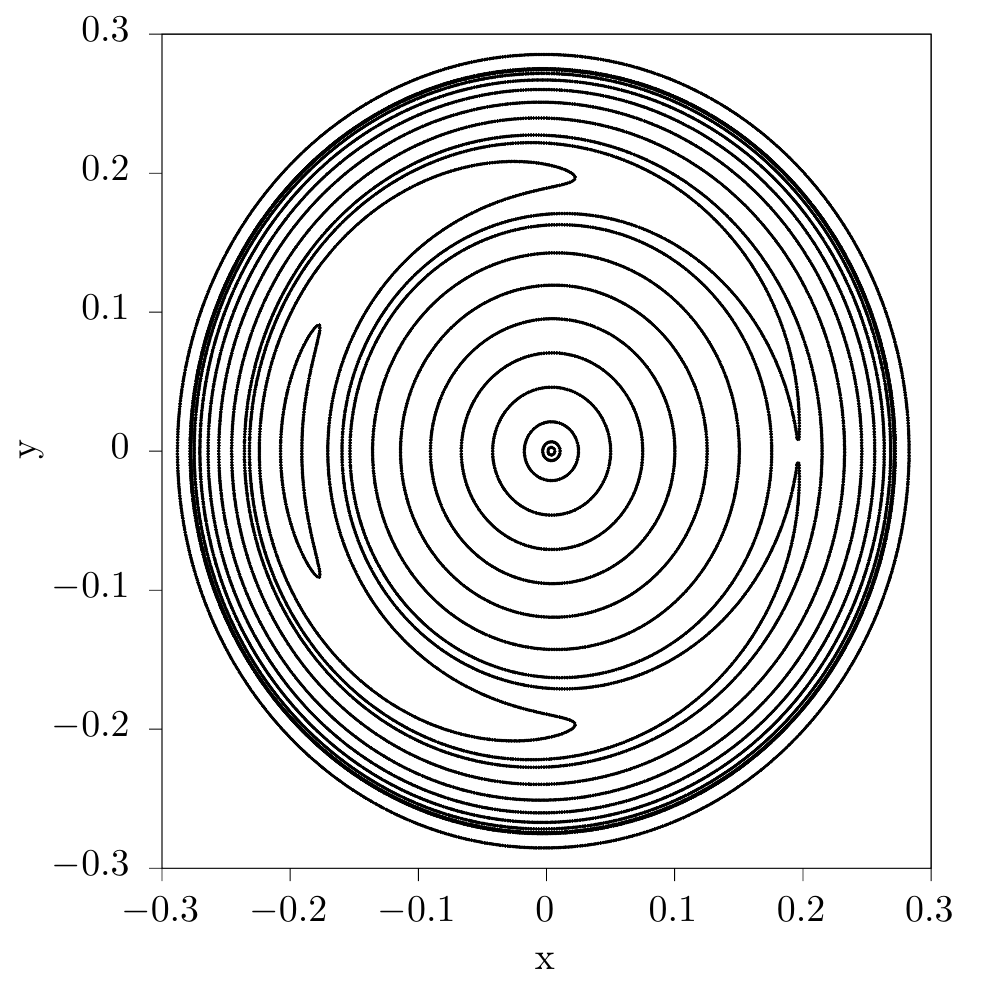}\\
\includegraphics[width=0.35 \columnwidth]{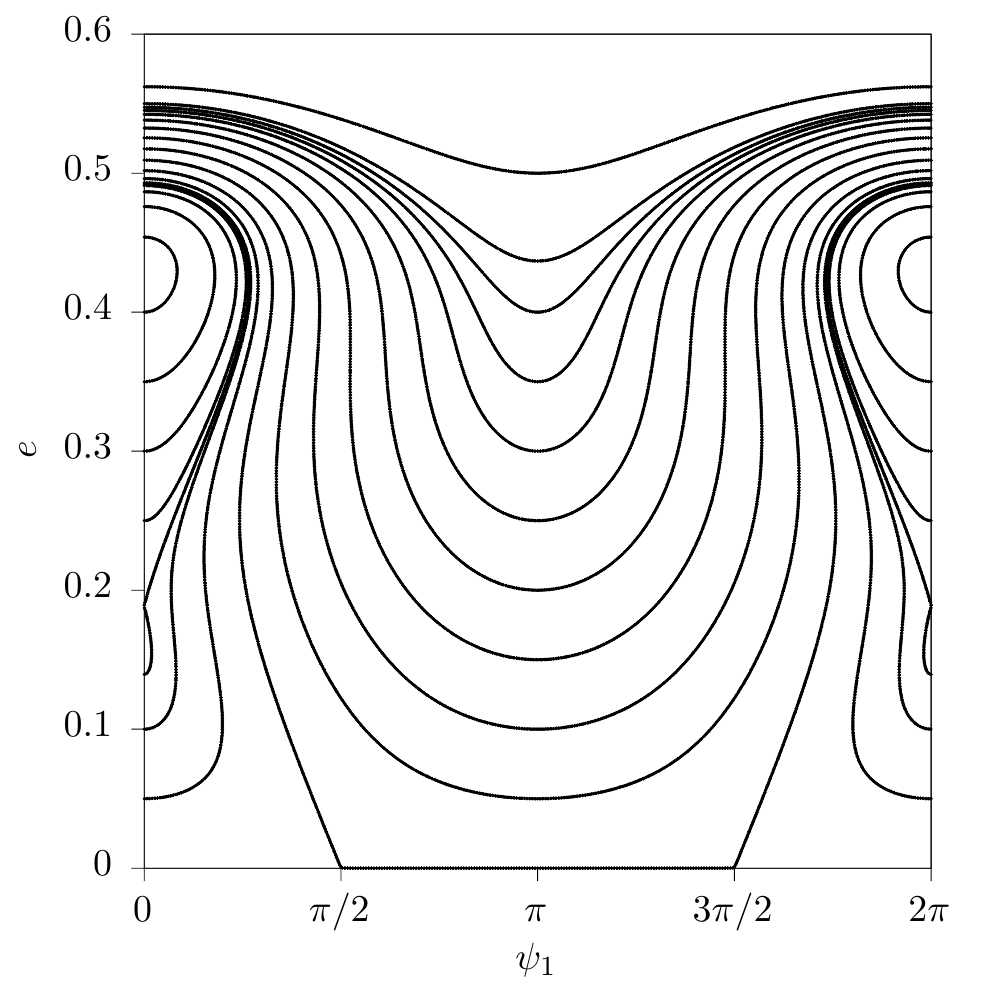}
\hspace{5pt}
\includegraphics[width=0.35 \columnwidth]{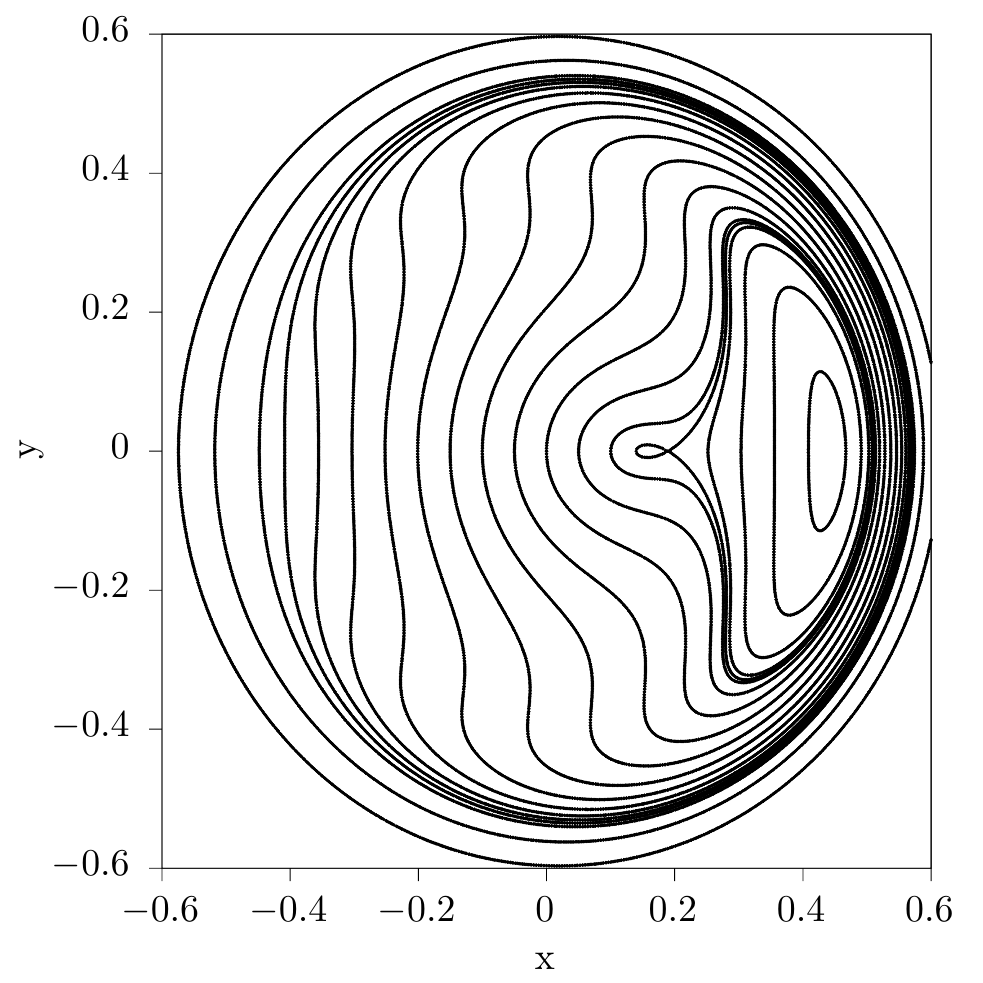}
\caption{Resonant phase space portraits for $\psi_1$, in $e-\psi$ and $x-y$ variables, according to the regions defined in Tab.~\ref{tab:res1eq} (top to bottom).}
\label{fig:phasespace}
\end{figure}

\noindent
For the resonant harmonic $\psi_1$ in Fig.~\ref{fig:bifplotpsi1} we present two bifurcation diagrams for $A/m=1$ m$^2/$kg (left) and $A/m=5$ m$^2/$kg (right). There are 5 distinct dynamical regions labelled with Latin numbers, defined in Tab.~\ref{tab:res1eq}, the corresponding phase spaces being shown in Fig.~\ref{fig:phasespace}.

\begin{table}
\centering
\begin{tabular}{ |c|c|c|c| } 
 \hline
 Region & total & $\psi_1=0$ & $\psi_1=\pi$  \\ 
 \hhline{|====|}
  I & 3 & 1 s & 1 s \& 1 u \\ 
 \hline
  II & 1 & 1 s & - \\ 
 \hline
 III & 5 & 2 s \& 1 u & 1 s \& 1 u \\
 \hline
  IV & 3 & 1 s \& 1 u & 1 s \\ 
 \hline
  V & 3 & 2 s \& 1 u & - \\ 
 \hline
\end{tabular}
\caption{Number of equilibria and their stability (s: stable, u: unstable) for the resonance with argument $\psi_1$, corresponding to the five regions of Fig.~\ref{fig:bifplotpsi1}. The corresponding phase space is depicted in Fig.~\ref{fig:phasespace}.}
\label{tab:res1eq}
\end{table}

\noindent
As just mentioned, the red line corresponds to the resonance location based on the $J_2$ rates. We observe that for the low $A/m=1$ m$^2/$kg value, this line almost coincides with the transition from region I to region III and from I to II. From Tab.~\ref{tab:res1eq} it is easy to see that the following bifurcations take place:

\smallskip
\noindent
$\psi=0$:
\begin{itemize}
\item transition from region I to region III: saddle-center; 
\item transition from region V to region II: saddle-center;
\end{itemize}
$\psi=\pi$:
\begin{itemize}
\item transition from region III to region V: saddle-center;
\item transition from region I to region II: saddle-center.
\end{itemize}

\begin{table}[htbp!]
\centering
\begin{tabular}{ |c|c|c|c| } 
 \hline
 Region & total & $\psi_2=0$ & $\psi_2=\pi$  \\ 
 \hhline{|====|}
  I & 1 & - & 1 s \ \\ 
  \hline
  II & 3 & 1 s \& 1 u & 1 s  \\ 
 \hline
\end{tabular}
\caption{Number of equilibria and their stability (s: stable, u: unstable) for the resonance with argument $\psi_2$, corresponding to the two regions of Fig.~\ref{fig:bifplotpsi2}.}
\label{tab:res2eq}
\end{table}

\begin{figure}[htbp!]
\centering
\includegraphics[width=0.9\columnwidth]{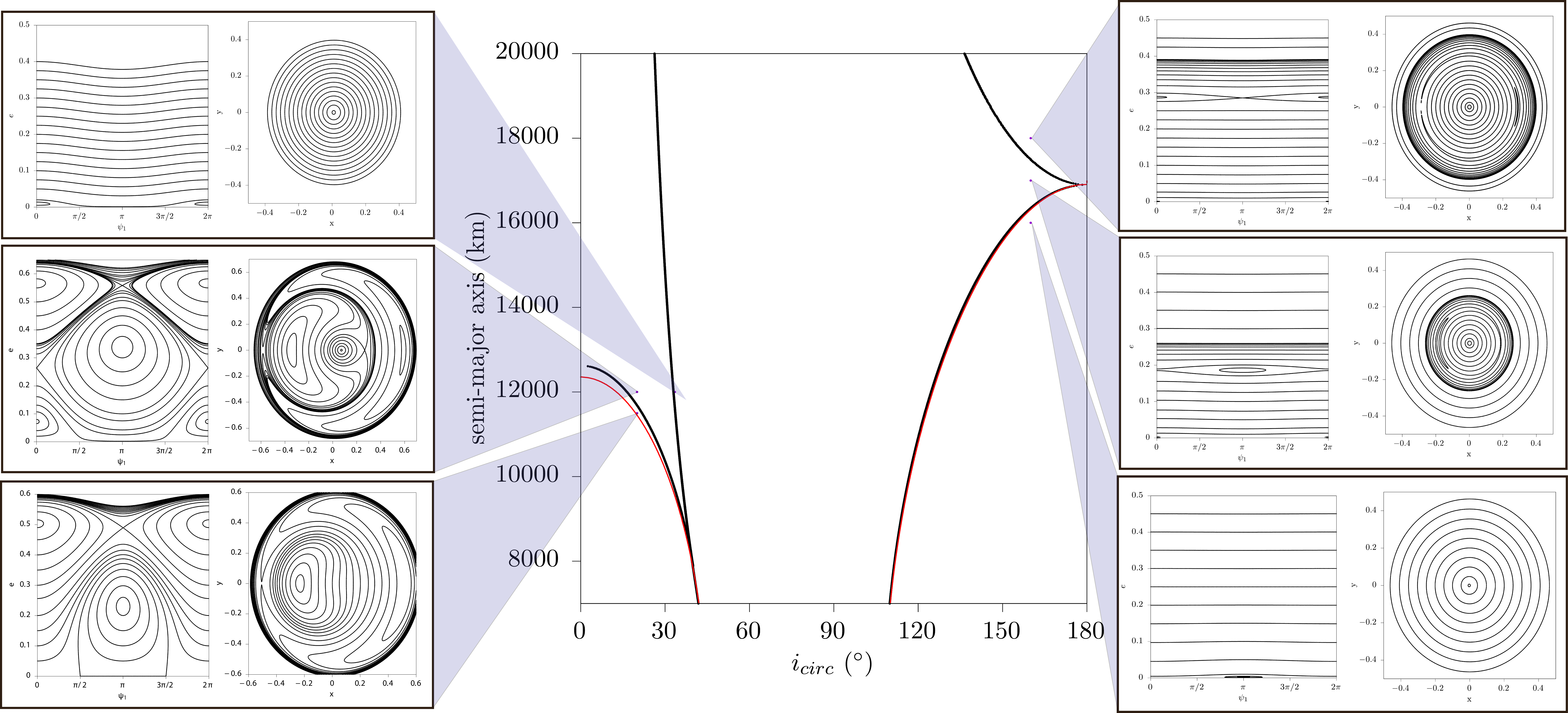} \\
\bigskip
\includegraphics[width=0.9\columnwidth]{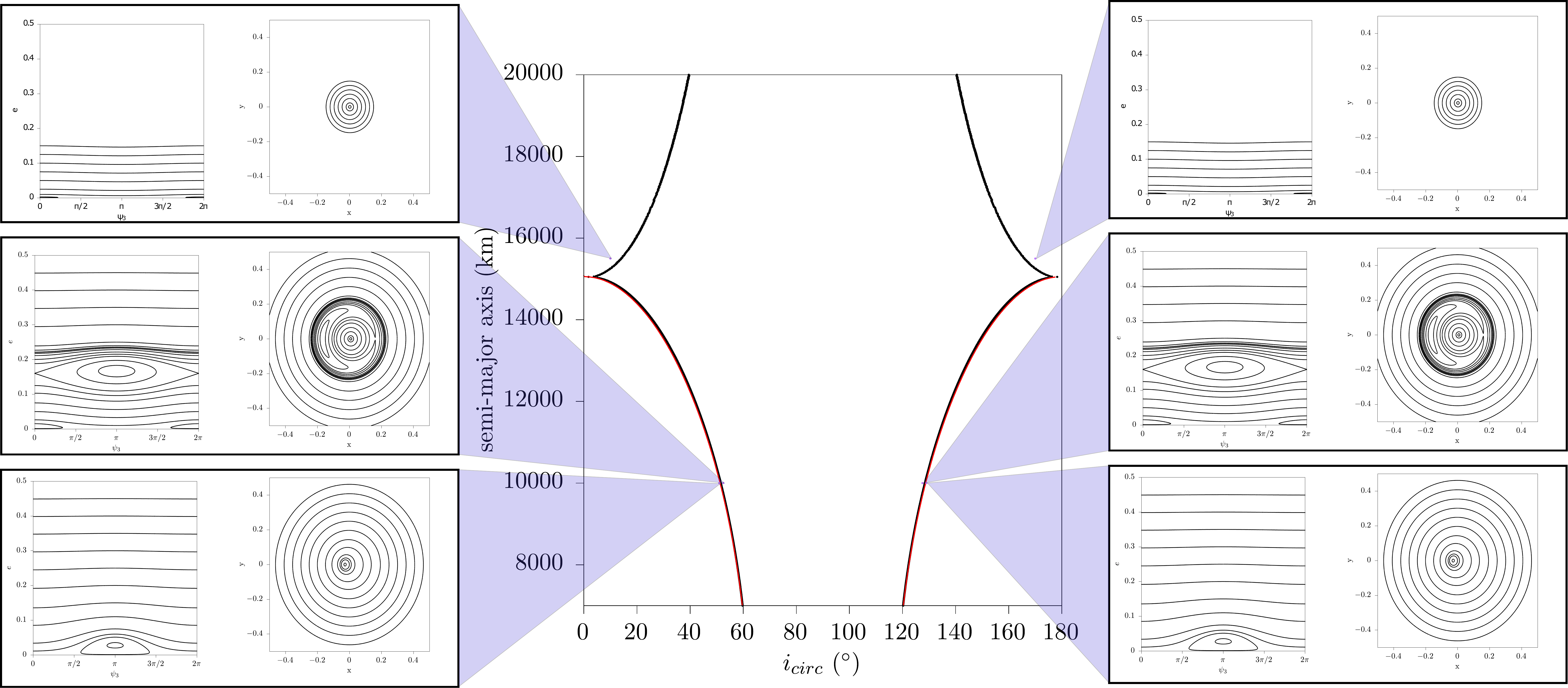} \\
\bigskip
\includegraphics[width=0.9\columnwidth]{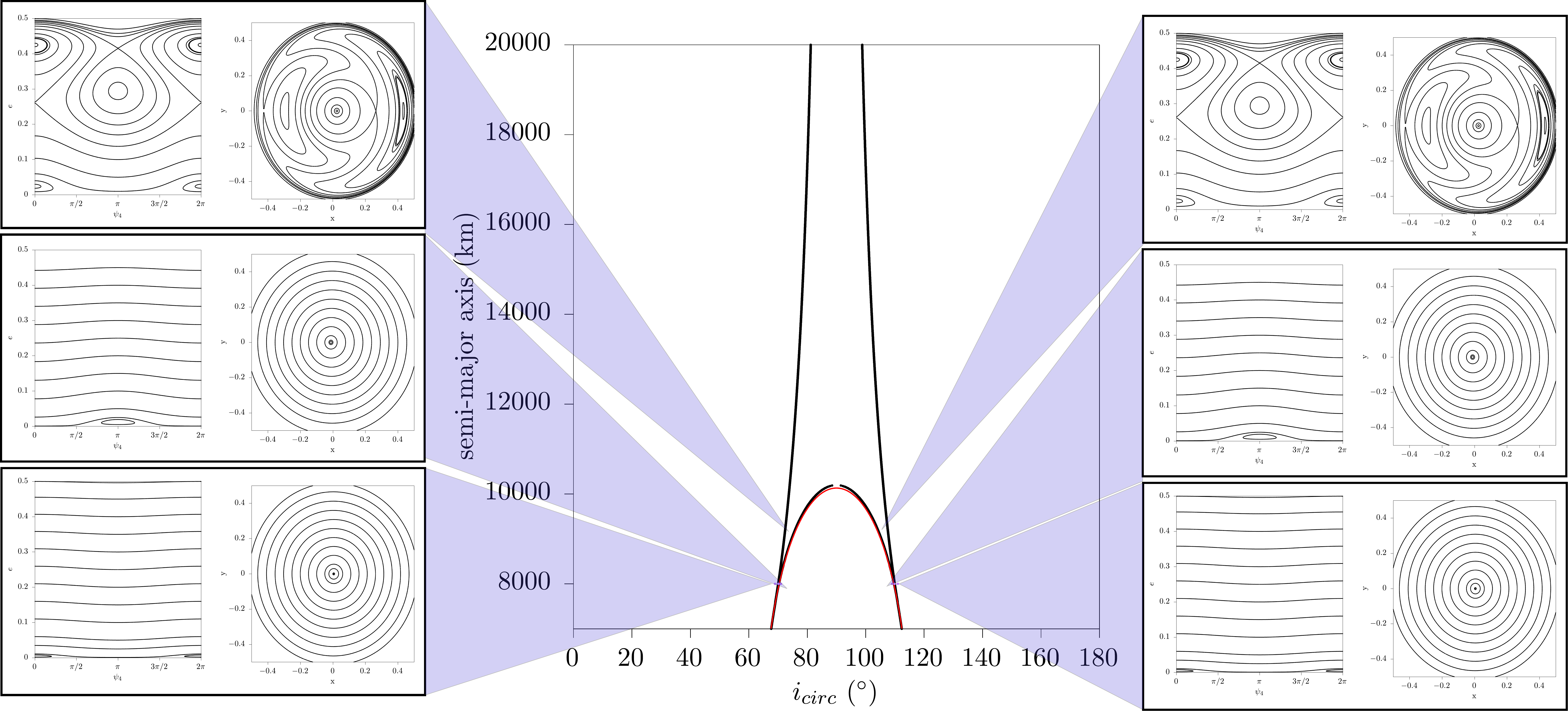}
\caption{The bifurcation diagram for the resonances with arguments  $\psi_1$ (top), $\psi_3$ (middle) and $\psi_4$ (bottom). The red curves correspond to the bifurcation that is obtained considering only the oblateness effect in $\dot\omega$ and $\dot\Omega$.}
\label{fig:bifurcation_134}
\end{figure}

\noindent
The effect of an increased $A/m$ on the bifurcation diagram is shown in the right panel of Fig.~\ref{fig:bifplotpsi1}. In this case, the exact position of the bifurcation is no longer predicted from the $J_2$ only rates as significant contributions are added to the $\dot{\omega}$ and $\dot{\Omega}$ rates due to SRP. The number of regions and their structure remains the same, however, the boarders between the different regions considerably alter. Specifically, for a fixed $i_{circ}$ the bifurcation for region I to regions II and III seem to happen at higher altitudes. This is in agreement to previous findings on the planar case ($i_{circ} = 0$) of $\psi_1$ \cite{LucCol2012}.

A similar analysis is now performed for the resonance with argument $\psi_2$. Assuming the same range semi-major axis as in $\psi_1$ case, we expect this harmonic to dominate in a range of inclinations from $i\in[76:84]$. The bifurcation diagram is shown in Fig.~\ref{fig:bifplotpsi2} and the equilibria with their associated stability are given in Tab.~\ref{tab:res2eq}.

\noindent
The situation in this case appears to be more straightforward. There are two distinct regions: in region I there exists only 1 stable equilibrium at $\psi_2=\pi$ while in region II two more equilibria, 1 stable and unstable, appear at $\psi_2 = 0$ after another saddle-center bifurcation. The position of the bifurcation for $A/m=1$ m$^2/$kg is approximately estimated from the $J_2$ resonant condition (red curve in Fig.~\ref{fig:bifplotpsi2}).

\begin{figure}
\centering
\includegraphics[width=0.45 \columnwidth]{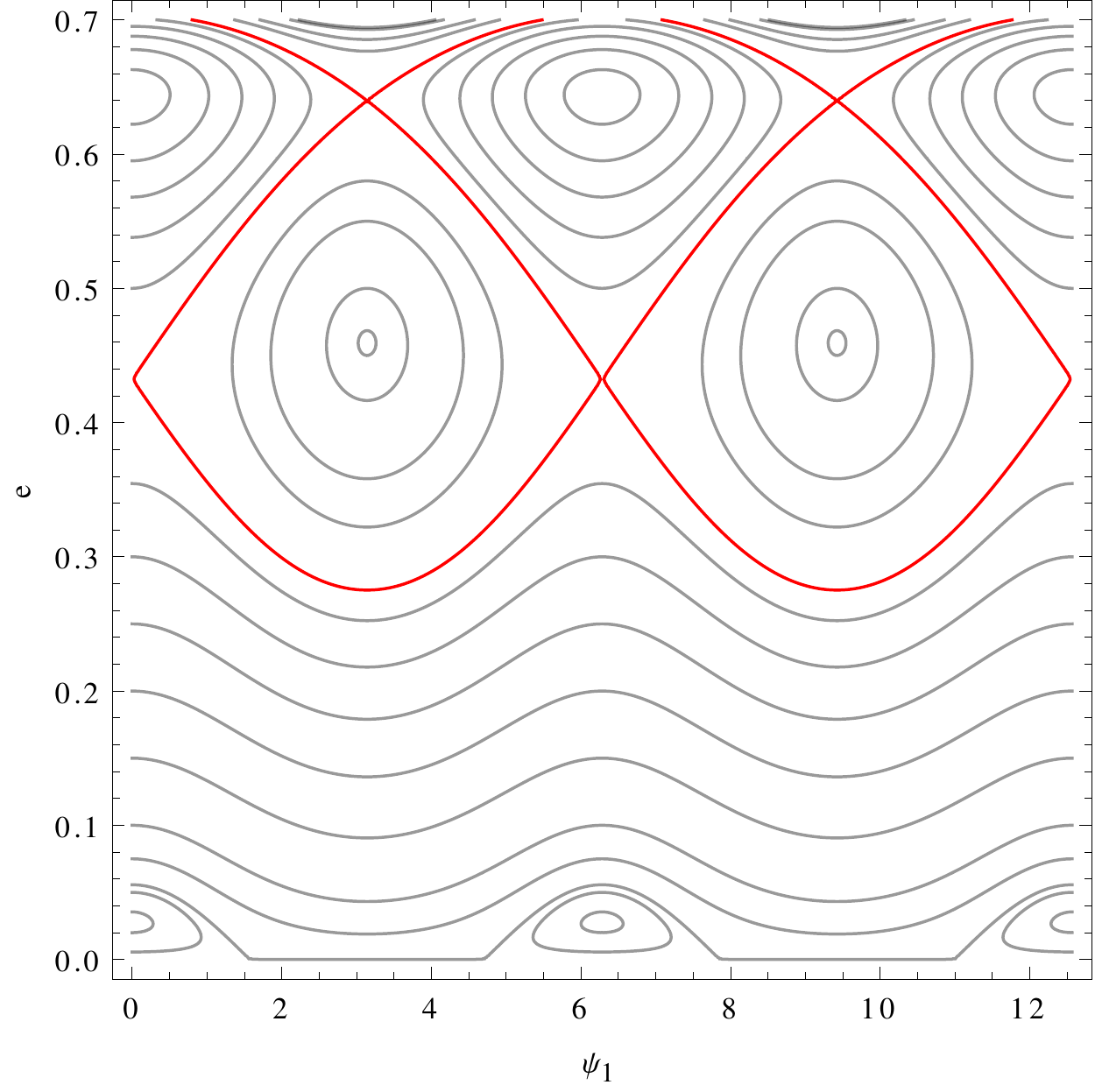} \hspace{0.2cm}\includegraphics[width=0.45 \columnwidth]{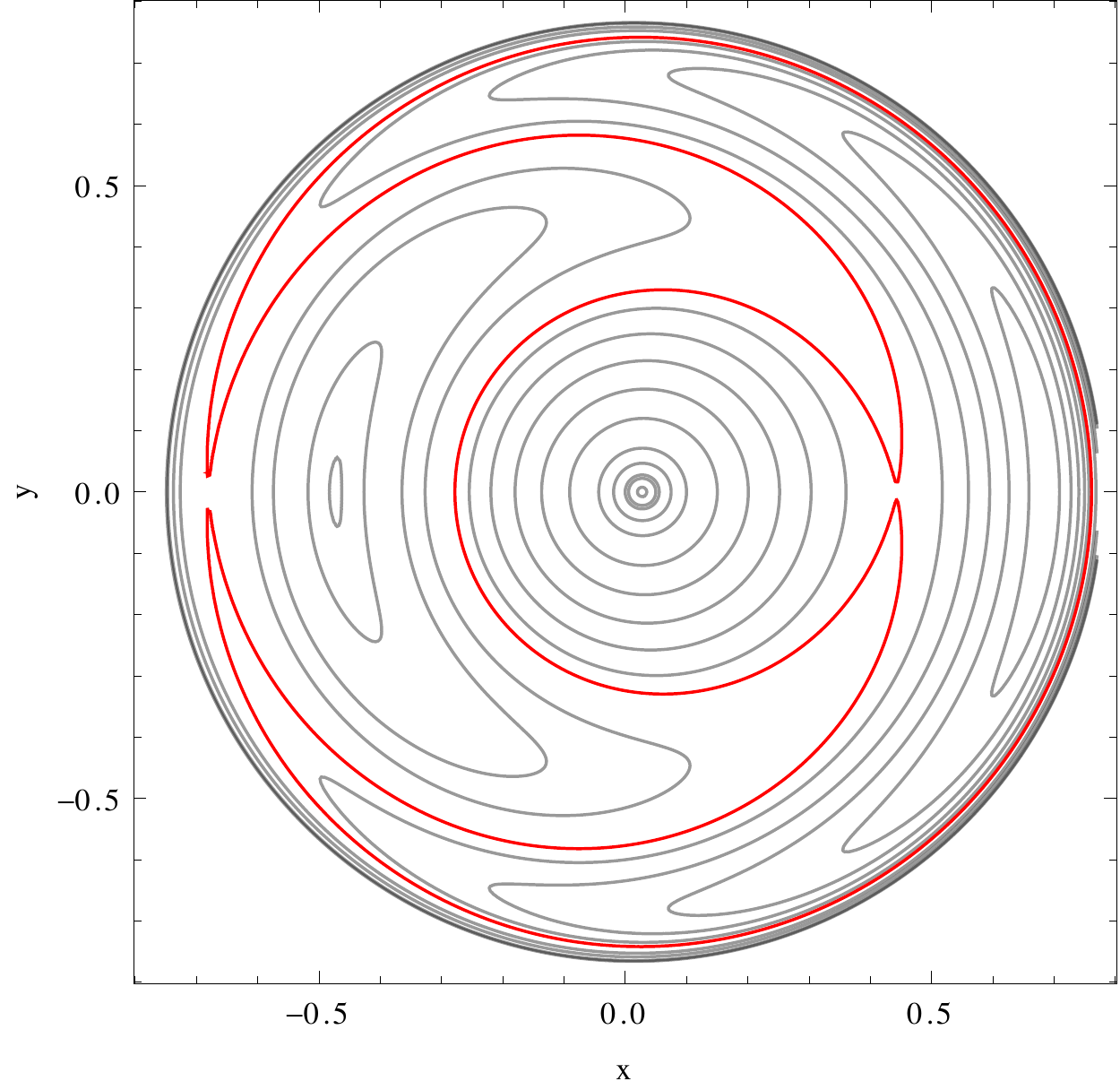}
\caption{Resonant phase space portrait for $\psi_1$, in $e-\psi$ and $x-y$ variables, in the case of a saddle connection (in red). On the left we extended the $\psi_1-$range to highlight the connection between the two saddle points.}
\label{fig:saddle_connection}
\end{figure}

Figure \ref{fig:bifurcation_134} extends the analysis by displaying the bifurcation diagrams for resonances $\psi_1$, $\psi_3$ and $\psi_4$, assuming $A/m=1$ m$^2/$kg and a range of semi-major axis  $a\in[7000:20000]$. As also found in \cite{ACR_CMDA2019}, the maximum number of equilibrium points is 5 also for $\psi_4$, and it is interesting to note that for $\psi_1$ it can occur another type of bifurcation apart from the saddle-center: the transcritical one beyond $a\sim 17000$ km and $i_{circ}\in[130^\circ:150^\circ]$. For the same resonant argument $\psi_1$, we can detect the saddle connection bifurcation, as described in \cite{Breiter2003}. An example is shown in Fig.~\ref{fig:saddle_connection}: in red it is highlighted the separatrix that connects two saddle points.

\noindent
The bifurcation diagram for $\psi_5$ and $\psi_2$ are symmetrical with respect to $i_{circ}=90^{\circ}$, the same holds for $\psi_6$ and $\psi_1$.

\section{Deorbiting configuration}\label{sec:deorbiting}

The phase space analysis can be exploited to compute the initial conditions, in terms of orbital elements and area-to-mass ratio, that can lead to an atmospheric reentry. Since the semi-major axis does not change under the dynamics considered, a natural deorbiting can take place only if the eccentricity increases as much as to attain the critical value  $e_{cr}=1- r_{\oplus}/a$. 

We consider the case where this condition occurs at either $\psi=0$ or $\psi=\pi$, meaning that the area-to-mass ratio required is the minimum. Moreover, mindful of the phase space behavior depicted in the previous section (see Fig.~\ref{fig:phasespace}), the steepest eccentricity growth from a circular orbit takes place starting from $\psi=\pi/2\lor 3\pi/2$ and following the stable direction associated with a hyperbolic equilibrium point or a libration curve associated with an elliptic equilibrium point. Based on this and following the idea presented in \cite{LCM_JSR2013} for the planar case, the initial conditions that can enable a natural deorbiting from a circular orbit in the spatial case can be obtained by solving the following equation
\begin{equation}\label{eq:lambda_ham}
\mathcal{H}_{\psi_j}(\Psi(e=0),\psi;L,\Pi) - \mathcal{H}_{\psi_j}(\Psi(e=e_{cr}),\psi;L,\Pi)=0, \quad \psi=0\lor \pi,
\end{equation} 
as a function of $C_{SRP}$ (or, equivalently, $A/m$). In particular,
\begin{equation}\label{eq:csrp}
\begin{aligned}
C_{SRP} =& \pm \frac{\mu }{6 L^{10} n_1^2 \mathcal{T}_j \eta_{cr}^5 e_{cr}}   \Big\{ 4 L^9 n_1 n_3 n_s (\eta_{cr}-1) \eta_{cr}^5\\
& - \mathcal{C}_{J_2} L^2 (n_1^2 - 3 n_2^2) \eta_{cr}^2 (\eta_{cr}^3-1) \mu^3 \\
& + 6 \mathcal{C}_{J_2} L n_2 \eta_{cr} (-1 + \eta_{cr}^4) \mu^3 \Pi +  3 \mathcal{C}_{J_2} ( \eta_{cr}^5-1) \mu^3 \Pi^2 \Big \}
\end{aligned}
\end{equation}
where $\eta_{cr} = \sqrt{1-e_{cr}^2}$, and the coefficients $T_j$ are computed using $c_i = \frac{L n_2 \eta_{cr} + \Pi}{L n_1 \eta_{cr}}$ and $s_i = \sqrt{1-c_i^2}$. The $\pm$  sign depends on whether the critical eccentricity is attained at $\psi=0$ or $\psi=\pi$, respectively. Note that only the positive solutions is admissible, because it represents a physical area-to-mass ratio. Moreover, it should be noticed that the solution does not provide any information on the time required to cover the invariant curve up to $e=e_{cr}$.

\noindent 
In the above expression, the only unknown is $\Pi$ that can be set considering that the deorbiting starts from $e_0=0$ and $\psi_0=\pi/2\lor 3\pi/2$, following a resonant curve. Given the value of semi-major axis $a$, the value of the inclination can be found by computing the resonant condition $\dot\psi=0$ at that configuration. Following \cite{ACR_CMDA2019}, the resonant condition $\dot\psi=0$ at $\psi=\pi/2\lor 3\pi/2$ does not depend on the solar radiation pressure, but only on the oblateness effect. In particular, $i_0$ can be obtained by solving the quadratic equation in $\cos i_0$ \cite{ACR_CMDA2019}
\begin{equation}\label{eq:res}
c_1\cos^2{i_0}+c_2\cos{i_0}+c_3=0,
\end{equation}
with
\begin{equation*}\label{eq:coef_res2}
c_1=\frac{15n_2J_2r^2_{\oplus}n}{4a^2(1-e^2)^2};\quad c_2=-\frac{3n_1J_2r^2_{\oplus}n}{2a^2(1-e^2)^2}; \quad c_3=n_3n_S-\frac{3n_2J_2r^2_{\oplus}n}{4a^2(1-e^2)^2}.
\end{equation*}

 \subsection{Results}
 
We have applied the procedure just described to the range $a\in[6978:15000]$ km, considering a discretization of $\Delta a=$ 10 km. The lower limit for the range in $a$ reflects the fact that below about 600 km of altitude a circular orbit decays naturally in 25 years due to the effect of the atmospheric drag.

\begin{figure}
\centering
\includegraphics[width=0.49\columnwidth]{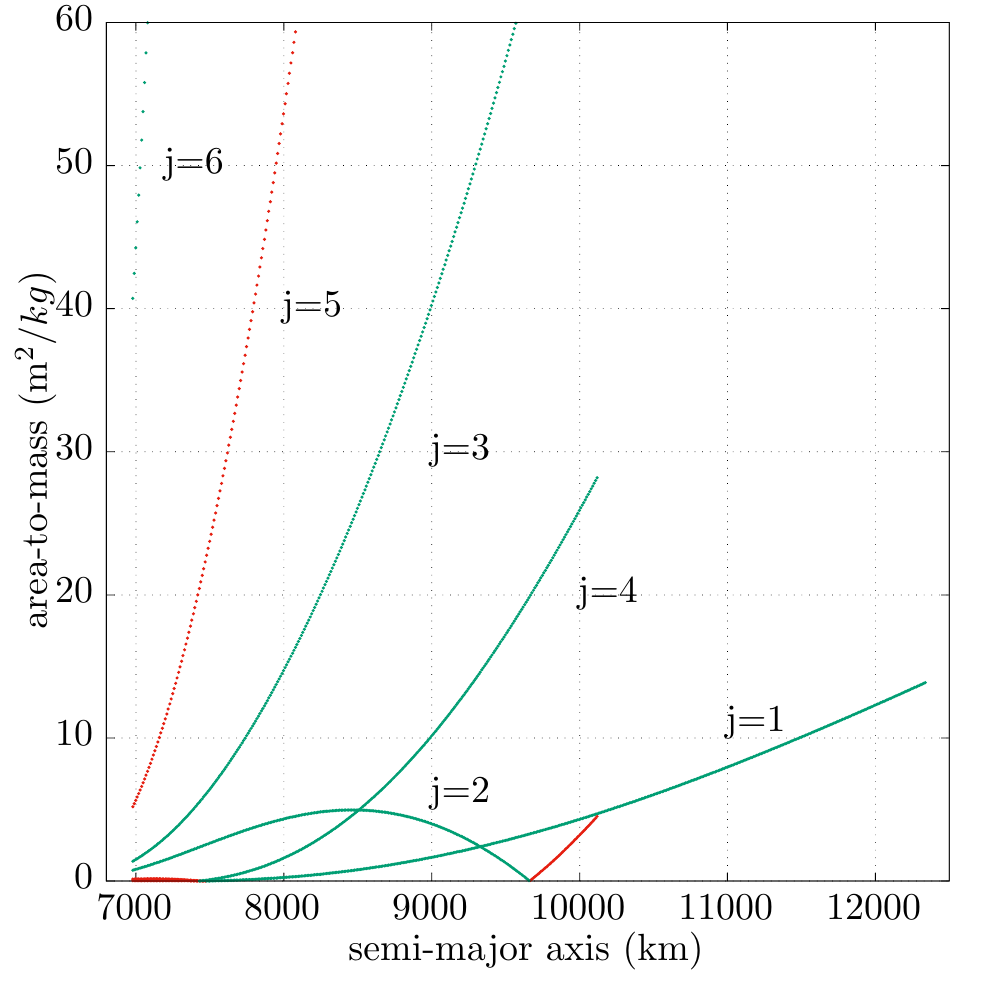} \includegraphics[width=0.49\columnwidth]{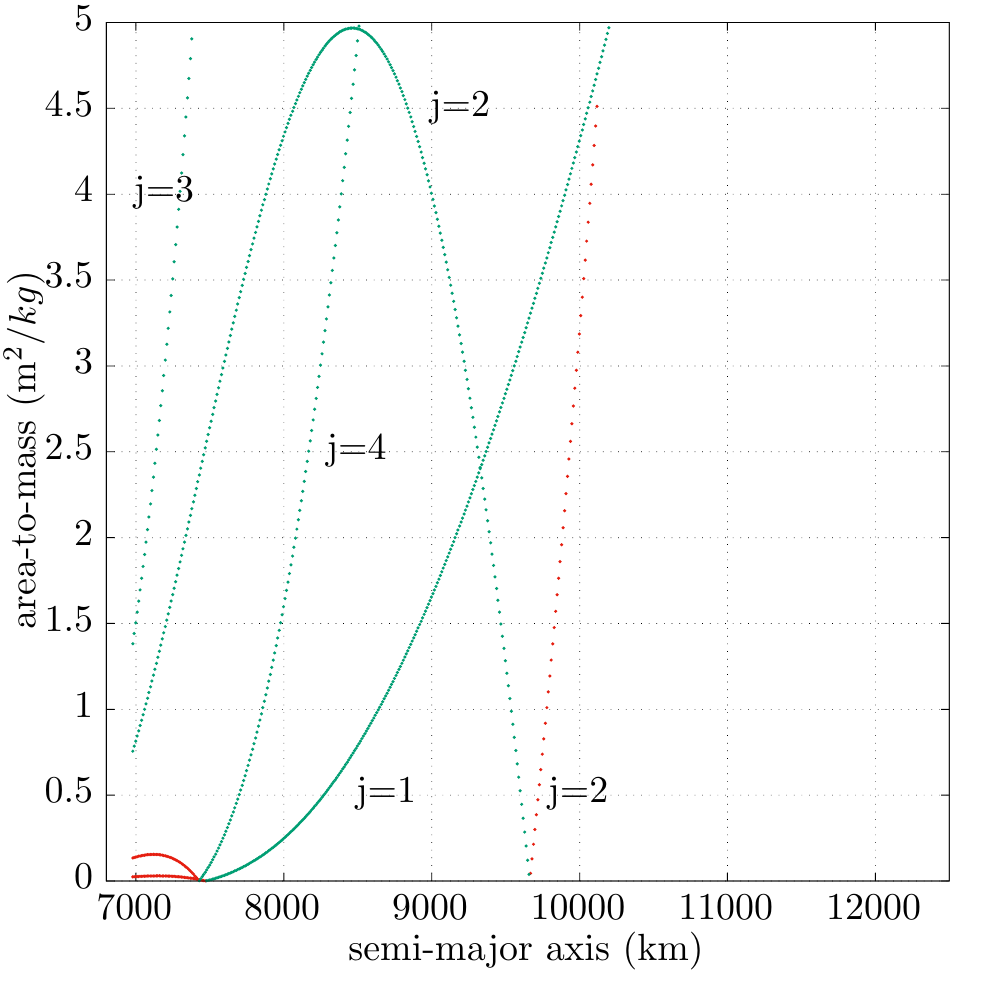}
\caption{Minimum area-to-mass ratio required to obtain a natural deorbiting, according to Eq.~(\ref{eq:csrp}). Red:  the deorbiting occurs at $\psi=0$; green: at $\psi=\pi$. Right: a closer view at feasible values of $A/m$.}
\label{fig:aa2m}
\end{figure}

In Fig.~\ref{fig:aa2m}, we show the area-to-mass ratio computed by means of Eq.~(\ref{eq:csrp}), as a function of initial semi-major axis for the six resonant arguments, for prograde orbits. The color reports if the deorbiting occurs at $\psi=0$ (red) or $\psi=\pi$ (green). It should be noticed that, as already detected numerically \cite{ASRV_CMDA2018,SASRS_ASR2019,Rossi2020}, the resonant terms $j=5,6$ are the least effective ones, requiring very high values of $A/m$. Moreover, the conditions depicted always correspond to a libration curve.

 \begin{figure}
\centering
\includegraphics[width=0.32\columnwidth]{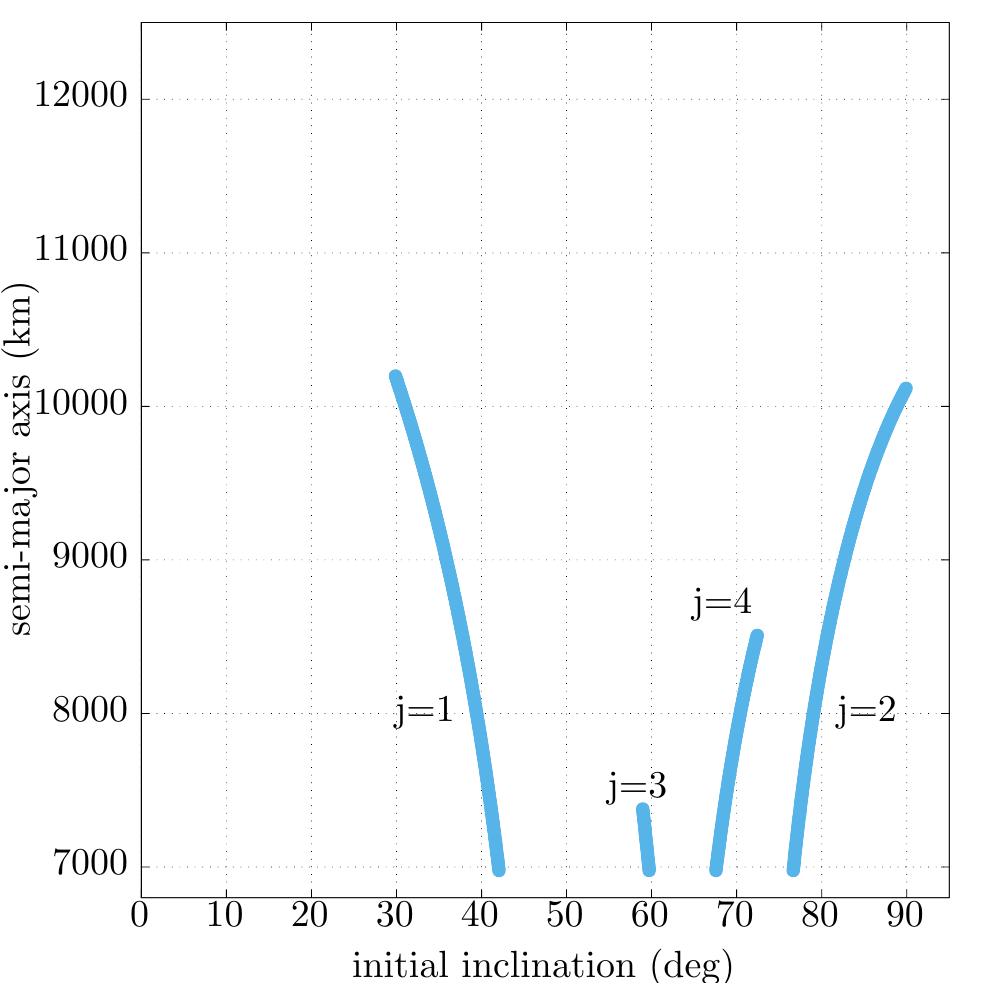} \includegraphics[width=0.32\columnwidth]{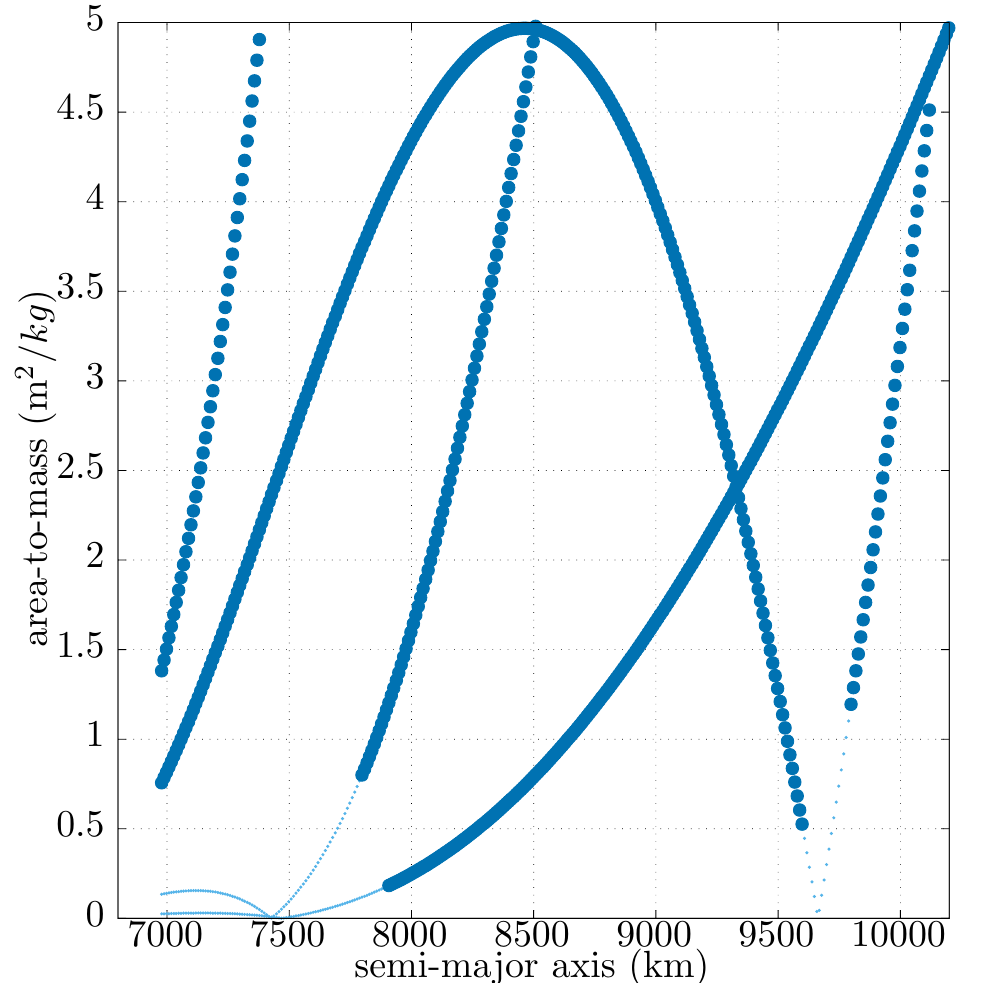} \includegraphics[width=0.32\columnwidth]{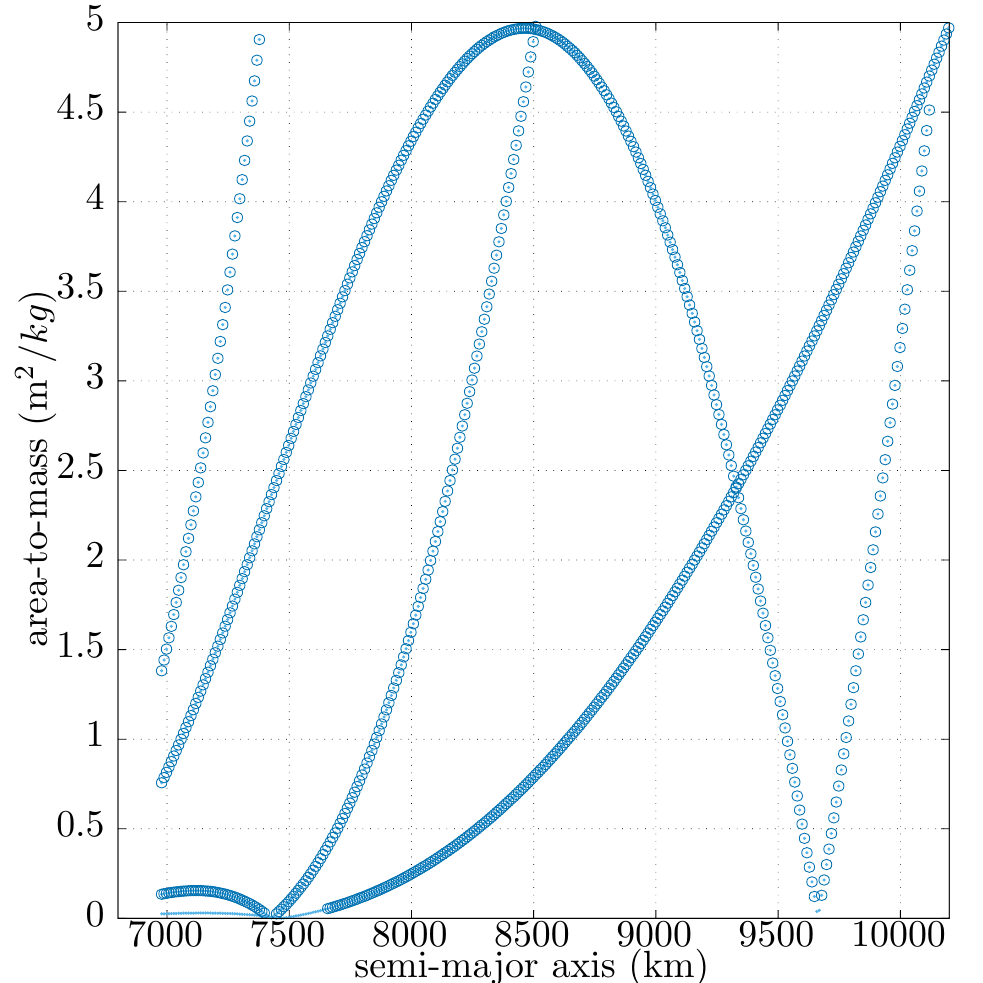}
\caption{Left: semi-major axis and inclination for the circular orbits corresponding to the minimum area-to-mass depicted in Fig.~\ref{fig:aa2m} on the right. Middle: the blue solid thicker points correspond to the conditions that can deorbit in less than 50 years. Right: the blue empty thicker points correspond to the conditions such that the approximation of neglecting the solar semi-secular terms is valid.}
\label{fig:aa2m_bis}
\end{figure}

The area-to-mass range $A/m\in[1:5]$ m$^2/$kg are characteristic of solar sails that are feasible with the current technology developments \cite{Colombo2018}. These are devices we can envisage to embed a spacecraft to deorbit them passively at the end-of-life. Higher values can represent clouds of fragments of space debris and can be used, for instance, to see if specific corridors are actually followed by fragments reentering the Earth after a collision. The option of setting specific observational campaigns can be explored in the future on the basis of the analysis presented here.

In Fig.~\ref{fig:aa2m_bis}, we show, on the left, the value of initial inclination and semi-major axis corresponding to the conditions depicted in Fig.~\ref{fig:aa2m} on the right. In the middle, we show with a thicker solid point the configurations that actually attain $e_{cr}$ in less than 50 years. Such limit can be considered conservative, because in the numerical propagation only the Earth's oblateness and the SRP are taken into account. The atmospheric drag, that will speed up the reentry for altitudes below 700 km (or higher, depending on $A/m$), is neglected. The cases where a 50-year reentry is not achieved are associated to two different behaviors:
\begin{itemize}
\item either the dynamics is too slow and this happens, in particular, after the cusps that can be noticed in the figure;
\item or there exist two curves, not connected, associated with the same value of $\mathcal{H}$, for a same integral of motion $\Lambda$. This happens, in particular, for $j=1$ and $j=4$ before for values of semi-major axis lower than $a=$ 7500 km.  
\end{itemize}
 In the same Fig.~\ref{fig:aa2m_bis}, on the right, we display with a thicker empty point the configurations such that the hypothesis of neglecting the semi-secular solar gravitational perturbations is valid (recall Sec.~\ref{sec:semisec}).  We consider that the approximation is valid if the ratio between the two effects is larger than 1000. It can be noticed that this is true in all the relevant cases shown in the middle panel.

\section{Conclusions}
\label{sec:concl}

In this work, we have extended the analytical development of the equations of motion associated with the coupled oblateness -- solar radiation pressure effect. In particular, the Hamiltonian formulation, only partially explained in \cite{Alessi_MNRAS,ACR_CMDA2019} has been here accurately presented. On this basis, the whole phase space in Earth orbit up to $a=20000$ km has been described by means of a dynamical taxonomy, that shows where the main bifurcations occurs, in particular the saddle-center, the transcritical and the saddle connection ones. Finally, the analytical value of the area-to-mass ratio required to de-orbit has been provided, for the first time in the three-dimensional case, and the corresponding applications for a feasible disposal strategy by means of a solar sail has been presented.

The model presented is a first approximation of how the solar radiation pressure effect coupled with the planetary oblateness can act on a satellite. The model neglects more complex issues, like shadows or short-term effects, but previous numerical works showed that the solutions shown persist when considering a full model, that includes the mentioned issues. This aspect was faced in particular in \cite{ASRV_CMDA2018,SASRS_ASR2019}.

The coupling with the solar gravitational perturbation that can occur at higher altitudes will be considered in the future.

\begin{acknowledgements}
This article is part of the COMPASS project, that has received funding from the European Research Council (ERC) under the European Union's Horizon 2020 research and innovation programme (Grant agreement No. 679086). The authors would like to acknowledge the previous studies cited in this paper performed in the framework of the European Commission Horizon 2020 ReDSHIFT project (Grant agreement No. 687500) and collaborations therein. Elisa Maria Alessi would like to acknowledge the support of CNR-IFAC and CNR-IMATI for the completion of this work. The authors would like to thank the anonymous reviewers for their useful suggestions.
\end{acknowledgements}

\section*{Compliances with ethical standards}
\textbf{Conflict of Interest Statement} On behalf of all authors, the corresponding author states that there is no conflict of interest.
%\nocite{*}
\bibliographystyle{spmpsci}      % mathematics and physical sciences
\bibliography{mybibfile}  % name your BibTeX data base

%\begin{thebibliography}{}
% etc
%\end{thebibliography}

\end{document}